\documentclass[10pt,a4paper]{article}
\usepackage[english]{babel}
\usepackage[latin1]{inputenc}
\usepackage{amsfonts,amsbsy,bm,euscript,mathrsfs}
\usepackage{amssymb,stmaryrd,faktor}
\usepackage[tbtags]{amsmath}
\usepackage{bbm}
\usepackage{graphicx}
\usepackage[title,titletoc]{appendix}
\usepackage[bookmarks=true,colorlinks=true,linkcolor=blue,citecolor=blue,urlcolor=blue,bookmarksnumbered]{hyperref}

\usepackage{dsfont}
\usepackage{collref}
\usepackage{lmodern}
\usepackage{mathrsfs}
\usepackage{mathtools}
\usepackage{bbm}
\usepackage{braket}
\usepackage{slashed}

\usepackage{graphicx}
\usepackage{booktabs}
\usepackage{subfig}
\usepackage{tikz}
\usetikzlibrary{plotmarks,calc,decorations,decorations.pathmorphing}

\numberwithin{equation}{section}

\makeatletter
\renewcommand\section{\@startsection {section}{1}{\z@}
{-3.5ex \@plus -1ex \@minus -.2ex}
{2.3ex \@plus.2ex}
{\normalfont\Large\bfseries}}
\renewcommand\subsection{\@startsection{subsection}{2}{\z@}
{-3.25ex\@plus -1ex \@minus -.2ex}
{1.5ex \@plus.2ex}
{\normalfont\large\bfseries}}
\makeatother

\def\a{\alpha}

\def\p{\phi}

\usepackage{blkarray}

\begin{document}


\thispagestyle{empty}
\begin{flushright}\footnotesize\ttfamily
DMUS-MP-26-03
\end{flushright}
\vspace{2em}

\begin{center}

{\Large\bf \vspace{0.2cm}
{\color{black} \large On the commuting charges and the Bethe ansatz for the quantised Kadomtsev-Petviashvili equation}} 
\vspace{1.5cm}

\textrm{Chiara Paletta$^{a}$ and Alessandro Torrielli$^{b}$ \footnote[1]{\textit{E-mail:} \texttt{chiara.paletta@fmf.uni-lj.si, a.torrielli@surrey.ac.uk}}}
\vspace{0.8cm}
\\
\vspace{0.3cm}
$^a$ \textit{Faculty of Mathematics and Physics, University of Ljubljana,
Jadranska ulica 19,
1000 Ljubljana}
\\
$^b$ \textit{School of of Mathematics and Physics, University of Surrey, Guildford, GU2 7XH, UK}
\vspace{0.3cm}

\end{center}

\vspace{2em}

\begin{abstract}\noindent 
 
We study the quantised version of the Kadomtsev-Petviashvili equation in the formulation proposed in \cite{Kozlowski:2016too}. We analyse two aspects. One is the conservation of the first higher charge $H_3$ (besides the total mass $H_0$, the momentum $H_1$ and the Hamiltonian $H_2$). We find the expression of $H_3$ and prove commutativity, provided we regularise a specific  ultraviolet divergence. The result then does not depend on the choice of a regulator, under fairly general assumptions. We then study the Bethe ansatz in finite volume and obtain a general solution to the two-body problem. We have used AI supervised by us to aid the proofs.

\end{abstract}

\newpage

\overfullrule=0pt
\parskip=2pt
\parindent=12pt
\headheight=0.0in \headsep=0.0in \topmargin=0.0in \oddsidemargin=0in

\vspace{-3cm}
\thispagestyle{empty}
\vspace{-1cm}

\tableofcontents

\setcounter{footnote}{0}

\section{Introduction}

The Kadomtsev-Petviashvili (KP) equation \cite{KP} (see \cite{KPweb} for a list of references) is one of the most famous integrable systems. It has the peculiarity of being a $2+1$-dimensional integrable system, relevant for a wide variety of physical effects. In the theory of water waves, it is used to model oceanic phenomena where the scale of the motion in the two-dimensions of the surface are much greater than the vertical depth of the waves. The equation then allows to model solitons (propagating waves retaining their shape and momentum through scattering against dispersion) which are either effectively one-dimensional, or genuinely bidimensional: these solitons are observed in the sea. The KP equation models also plasma, and waves in ferromagnetic media, as well as it plays a role in the theory of the Bose-Einstein condensate. It is well-known that it can be treated using the inverse-scattering transform, and for its nature it is a wonderful laboratory where to study integrability i) in higher dimensions and ii) in the absence of particle-number conservation, two areas which are recently becoming very relevant in the literature \cite{Benini,Borsato}. 

Whereas traditionally the KP equation has been studied as a classical integrable system, in \cite{Kozlowski:2016too} a proposal was advanced for its quantum version. The strategy of \cite{Kozlowski:2016too} has been of compactifying one of the directions of the surface of propagation into a circle, expanded in what we could call Kaluza-Klein modes of the reduced model, which in this fashion becomes again a $1+1$-dimensional model but with an infinite number of particles of growing integer mass. Despite the difficulties, integrability organises all these infinitely many particles according to conserved charges and allows a Bethe ansatz to be formulated for the exact solution of the problem. As always, the solution of the two-body mass-$1$ problem - plus a contribution from a single particle of mass $2$ - recursively determines all other sectors, which are organised according to the total mass. Remarkably, the exact $S$-matrix is determined in \cite{Kozlowski:2016too} and it has very interesting features potentially connected to quantum algebras \cite{Yangian}. 

The thesis \cite{Chiara} contains an analysis of the singularities of the two-body $S$-matrix with the aim of defining the bound-state spectrum, and a perturbative study of the scattering theory. The thesis is one of the first documents where the quantised version of KP proposed in \cite{Kozlowski:2016too} has been pursued, and it has served the purposes to demonstrating certain consistency requisite of the model. 

In this present paper, we take this study further and concern ourselves with the higher commuting charges - something which has not been tackled before and notoriously difficult also for simpler systems, such as the non-linear Schr\"odinger  equation \cite{diff,Davi,nls}. We then study the Bethe ansatz in finite volume - the Bethe equations - and find a remarkable combinatorial solution for the two-body eigenstate. 

The article is organised as follows. In section \ref{chargeH3sec}, we present the higher charge $H_3$ and in \ref{detailscommutation} we provide full details of the commutation relations. The proof is assisted by AI, and it has been completely cross-checked by the authors. In section \ref{betheansatzsection}, we analyse the Bethe ansatz and provide solution to the two-body problem. We first obtain perturbative solutions to the Bethe equations as expansions in the parameter $\beta$, and then use AI to resum these expressions. All AI-assisted steps have been fully cross-checked by the authors. We conclude the article in section \ref{conclusions}. Further technical details  on the commutation relations of the charges $H_0, H_1$ and $H_2$ are collected in appendix \ref{AppendixcommutatorsH0H1H2}.

\section{New charge $H_3$}
\label{chargeH3sec}
\subsection{Expression of the quantum charge $H_3$}
\label{chargeH3secexpression}
The classical charge $h_3$ is
\begin{align}
    &h_3(\sigma,x)=\frac{1}{2}
\left(\partial_\sigma^{-3}\varphi\right)
\left(\partial_x^3\varphi\right)
-\beta\,\varphi^2\,
\partial_\sigma^{-1}\partial_x\varphi
-\frac{3\gamma}{2}
\left(\partial_\sigma\varphi\right)
\left(\partial_x\varphi\right)
\label{expressionh3},
\end{align}
where $\varphi=\varphi(\sigma,x)$ is the classical field entering the KP equation, with $\sigma$ denoting the compact direction and $x$ the spatial coordinate.
$h_3$ is derived in this way:

\begin{itemize}
    \item $\frac{1}{2}
\left(\partial_\sigma^{-3}\varphi\right)
\left(\partial_x^3\varphi\right)$ comes from the expression (2.5) in \cite{Kozlowski:2016too};
\item $-\beta\,\varphi^2\,
\partial_\sigma^{-1}\partial_x\varphi-\frac{3\gamma}{2}
\left(\partial_\sigma\varphi\right)
\left(\partial_x\varphi\right)$ are calculated such that $\{H_3,B\}=-3 H_2$, where $B$ is the Galilei boost operator $B=\int _0^{2\pi} \frac{d \sigma}{2 \pi}\int_{-\infty}^{\infty}dx\,x \,h_0(\sigma,x)$, and $h_0=\varphi^2/2.$
\end{itemize}
In principle, this does not uniquely fix the charge since there may be elements in the kernel of the Galilei boost operator. 
In this case, we checked that the density $h_3(\sigma,x)$ of \eqref{expressionh3} is a good choice since $\{H_3,H_0\}=\{H_3,H_1\}=\{H_3,H_2\}=0$.

Now, by following the steps of \cite{Kozlowski:2016too}, we want to calculate the quantum version of $h_3$ and check that, by using the correspondence principle $[\cdot,\cdot]\to i \hbar\{\cdot,\cdot\}$, 
 \begin{align}
     &[H_3,H_0]=[H_3,H_1]=[H_3,H_2]=0
 \end{align}
 and  \begin{align}
     &[H_3,\mathcal{B}]=-3 i H_2,
     \label{H3Bquantum}
 \end{align}
where now $\mathcal{B}=\sum_{m\ge1}m\int dx \,x\,\Psi_m^\dagger\Psi_m$ and we set $\hbar=1$.
 
 However, if we proceed in the standard form of \cite{Kozlowski:2016too}, we obtain $H_3$ 

\begin{align}
    & \begin{aligned}
H_3
={}&
i\sum_{m\geq 1}\frac{1}{m^2}
\int_{-\infty}^{+\infty}dx\,
\Psi_m^\dagger\,\partial_x^3\Psi_m
\\
&+i
\sum_{m_1,m_2\geq 1}
\beta_{m_1m_2}
\int_{-\infty}^{+\infty}dx\,
\Bigg[
\frac{1}{m_1+m_2}
\bigl(\partial_x\Psi_{m_1+m_2}^\dagger\bigr)
\Psi_{m_1}\Psi_{m_2}
\\
&\hspace{4.2cm}
-\frac{2}{m_1}
\Psi_{m_1+m_2}^\dagger
\bigl(\partial_x\Psi_{m_1}\bigr)\Psi_{m_2}
-\frac{2}{m_2}
\Psi_{m_1+m_2}^\dagger
\Psi_{m_1}\bigl(\partial_x\Psi_{m_2}\bigr)
\\
&\hspace{4.2cm}
-\frac{1}{m_1+m_2}
\Psi_{m_1}^\dagger\Psi_{m_2}^\dagger
\bigl(\partial_x\Psi_{m_1+m_2}\bigr)
\\
&\hspace{4.2cm}
+\frac{2}{m_1}
\bigl(\partial_x\Psi_{m_1}^\dagger\bigr)
\Psi_{m_2}^\dagger\Psi_{m_1+m_2}
+\frac{2}{m_2}
\Psi_{m_1}^\dagger
\bigl(\partial_x\Psi_{m_2}^\dagger\bigr)
\Psi_{m_1+m_2}
\Bigg]\\
&-3i\gamma\sum_{m\geq 1}m^2
\int_{-\infty}^{+\infty}dx\,
\Psi_m^\dagger\,\partial_x\Psi_m,
\end{aligned}
\label{chargeH3}
\end{align}
where $\beta_{m_1,m_2}=\beta \sqrt{(m_1+m_2)m_1m_2}$ and $\Psi=\Psi(x)$.
The problem is that this $H_3$ does not satisfy \eqref{H3Bquantum}.

We found that the $\beta$-dependent term is the problematic one. One way to remedy this is by leaving generic constants in front of the coefficients $\frac{1}{m_1+m_2},\frac{1}{m_1},\frac{1}{m_2}$ of \eqref{chargeH3}. By imposing the commutation of the charges by brute force, we fix the constants and we get that $H_3$ should be as follows (notice that before we had a factor of 2, now it is 1): 
\begin{align}
\begin{aligned}
H_3
=&\;
i\sum_{m\geq 1}\frac{1}{m^2}
\int dx\,
\Psi_m^\dagger\,\partial_x^3\Psi_m
\\
&+
i
\sum_{m_1,m_2\geq 1}
\beta_{m_1, m_2}
\int dx\,
\Bigg[
\frac{1}{m_1+m_2}
\big(\partial_x\Psi_{m_1+m_2}^\dagger\big)
\Psi_{m_1}\Psi_{m_2}
\\
&\hspace{3.0cm}
-
\frac{1}{m_1}
\Psi_{m_1+m_2}^\dagger
\big(\partial_x\Psi_{m_1}\big)
\Psi_{m_2}
-
\frac{1}{m_2}
\Psi_{m_1+m_2}^\dagger
\Psi_{m_1}
\big(\partial_x\Psi_{m_2}\big)
\\
&\hspace{3.0cm}
-
\frac{1}{m_1+m_2}
\Psi_{m_1}^\dagger\Psi_{m_2}^\dagger
\big(\partial_x\Psi_{m_1+m_2}\big)
\\
&\hspace{3.0cm}
+
\frac{1}{m_1}
\big(\partial_x\Psi_{m_1}^\dagger\big)
\Psi_{m_2}^\dagger\Psi_{m_1+m_2}
+
\frac{1}{m_2}
\Psi_{m_1}^\dagger
\big(\partial_x\Psi_{m_2}^\dagger\big)
\Psi_{m_1+m_2}
\Bigg]\\
&-
3i\gamma
\sum_{m\geq 1}m^2
\int dx\,
\Psi_m^\dagger\,\partial_x\Psi_m.\label{newH3}
\end{aligned}
\end{align}

A similar phenomenon was also analyzed for the non-linear Schr\"odinger model in \cite{nls}.

As we will demonstrate, with this choice of constants we can achieve commutation of the higher charge $H_3$ provided a regulator is chosen to regularise an ultraviolet singularity. We have necessarily  to introduce such regulator, although it can be done in  a fairly general way without spoiling the result. This Hamiltonian satisfies \eqref{H3Bquantum}.

In what follows, we prove the vanishing of $[H_3,H_0]$, $[H_3,H_1]$, and $[H_3,H_2]$. For completeness, in Appendix \ref{AppendixcommutatorsH0H1H2} we also verify that $[H_0,H_1]$, $[H_0,H_2]$, and $[H_1,H_2]$ vanish.

We recall the definition of $H_0, H_1, H_2$

\begin{align}
    &H_0=
\sum_{m\in\mathbb{N}}
m
\int_{-\infty}^{+\infty} dx\,
\Psi_m^\dagger(x)\Psi_m(x).
\end{align}
Since
\begin{align}
    &[H_0,\Psi_m(x)]
=
-m\,\Psi_m(x),
&[H_0,\Psi_m^\dagger(x)]
=
m\,\Psi_m^\dagger(x),
\label{H0PsiandH0Psidag}
\end{align}
$H_0$ is the total mass.
\begin{align}
    &H_1=
-i
\sum_{m\in\mathbb{N}}
\int_{-\infty}^{+\infty} dx\,
\Psi_m^\dagger(x)\partial_x\Psi_m(x),
\end{align}

$H_1$ satisfies
\begin{align}
    &[H_1,\Psi_m(x)]
=
i\,\partial_x\Psi_m(x),
    &[H_1,\Psi_m^\dagger(x)]
=
i\,\partial_x\Psi_m^\dagger(x),
\label{comH1PsiandPsidag}
\end{align}

 $H_1$ is the generator of translation in $x$.
\begin{align}
\begin{aligned}
H_2
=&
-\sum_{m\geq 1}\frac{1}{m}
\int dx\,
\Psi_m^\dagger(x)\partial_x^2\Psi_m(x)
+
\sum_{m\geq 1}\gamma_m
\int dx\,
\Psi_m^\dagger(x)\Psi_m(x)
\\
&+
\sum_{m_1,m_2\geq 1}
\beta_{m_1,m_2}
\int dx\,
\left(
\Psi_M^\dagger(x)\Psi_{m_1}(x)\Psi_{m_2}(x)
+
\Psi_{m_1}^\dagger(x)\Psi_{m_2}^\dagger(x)\Psi_M(x)
\right),
\end{aligned}
\end{align}
$H_2$ is the Hamiltonian of a non-relativistic, Galilei-invariant system of one-dimensional Bose-particles, where the cubic term $\beta$ describe processes where 2 particles of masses $m_1$ and $m_2$ merge into one of mass $m_1+m_2$ and the respective splitting.

In Appendix \ref{AppendixcommutatorsH0H1H2}, we report the calculation of \cite{Kozlowski:2016too}, for the commutations between $H_0,H_1,H_2$.

\subsection{Commutation with $H_3$}
\label{detailscommutation}
\subsubsection{$[H_0,H_3]$} 
Since \(H_3\) is the sum of quadratic and cubic terms, we check them separately.

First, consider the quadratic terms
\begin{align}
\Psi_m^\dagger(x)\partial_x^3\Psi_m(x),
\qquad
\Psi_m^\dagger(x)\partial_x\Psi_m(x).
\label{quadratic}    
\end{align}
For any \(k\geq 0\), due to \eqref{H0PsiandH0Psidag}
\begin{align}
[H_0,\partial_x^k\Psi_m(x)]
=
-m\,\partial_x^k\Psi_m(x),    
\end{align}
and hence
\begin{align}
    \begin{aligned}
[H_0,\Psi_m^\dagger(x)\partial_x^k\Psi_m(x)]
&=
m\,\Psi_m^\dagger(x)\partial_x^k\Psi_m(x)
-
m\,\Psi_m^\dagger(x)\partial_x^k\Psi_m(x)
=0.
\end{aligned}
\end{align}

Thus the quadratic part of \(H_3\) commutes with \(H_0\).

Now consider the cubic terms.

We consider the joining and splitting terms,
\begin{align}
&    J_{m_1,m_2}(x)
=
\Psi_M^\dagger(x)\Psi_{m_1}(x)\Psi_{m_2}(x),
\label{Jjoining}\\
&J_{m_1,m_2}^\dagger(x)
=
\Psi_{m_1}^\dagger(x)\Psi_{m_2}^\dagger(x)\Psi_M(x).\label{Jsplitting}
\end{align}
For them,
\begin{align}    
&[H_0,J_{m_1,m_2}(x)]
=
\left(M-m_1-m_2\right)
J_{m_1,m_2}(x)
=0,\label{commutH0J}\\
&[H_0,J_{m_1,m_2}^\dagger(x)]
=
\left(m_1+m_2-M\right)
J_{m_1,m_2}^\dagger(x)
=0,\label{commutH0Jdag}
\end{align}
because \(M=m_1+m_2\).

The same argument applies if the derivative acts on any of the fields. For example,
\begin{align}
[H_0,\Psi_M^\dagger(x)\Psi_{m_1}(x)\partial_x\Psi_{m_2}(x)]
&=
\left(M-m_1-m_2\right)
\Psi_M^\dagger(x)\Psi_{m_1}(x)\partial_x\Psi_{m_2}(x)
=0.
\end{align}

Therefore every term in \(H_3\) commutes with \(H_0\). Hence
\begin{align}
[H_0,H_3]=0.    
\end{align}

\subsubsection{$[H_3,H_1]$} 
For every \(k\geq 0\), using \eqref{comH1PsiandPsidag}
\begin{align}
    [H_1,\partial_x^k\Psi_m(x)]
=
i\,\partial_x^{k+1}\Psi_m(x),
\qquad
[H_1,\partial_x^k\Psi_m^\dagger(x)]
=
i\,\partial_x^{k+1}\Psi_m^\dagger(x),
\label{comH1derpsi}
\end{align}

we write
\begin{align}
H_3=\int dx\,h_3(x),    
\end{align}

where \(h_3(x)\) is the local density of \eqref{newH3} built from the fields and their \(x\)-derivatives, with no explicit \(x\)-dependence.

First, consider the quadratic terms of \(H_3\), \eqref{quadratic}.

For any \(k\geq 0\),
\begin{align}
[H_1,\Psi_m^\dagger(x)\partial_x^k\Psi_m(x)]
&=
i(\partial_x\Psi_m^\dagger(x))\partial_x^k\Psi_m(x)
+
i\Psi_m^\dagger(x)\partial_x^{k+1}\Psi_m(x)
=
i\partial_x
\left(
\Psi_m^\dagger(x)\partial_x^k\Psi_m(x)
\right).
\end{align}

Now consider a typical cubic joining term, for example
\[
T_{m_1,m_2}(x)
=
(\partial_x\Psi_M^\dagger(x))\Psi_{m_1}(x)\Psi_{m_2}(x),
\qquad
M=m_1+m_2.
\]
Then, using \eqref{comH1derpsi}
\begin{align}    
\begin{aligned}
[H_1,T_{m_1,m_2}(x)]
&=
i(\partial_x^2\Psi_M^\dagger(x))\Psi_{m_1}(x)\Psi_{m_2}(x)
+i(\partial_x\Psi_M^\dagger(x))(\partial_x\Psi_{m_1}(x))\Psi_{m_2}(x)
\\
&\quad
+i(\partial_x\Psi_M^\dagger(x))\Psi_{m_1}(x)(\partial_x\Psi_{m_2}(x))
=
i\partial_x T_{m_1,m_2}(x).
\end{aligned}
\end{align}

The same argument applies to all the other cubic terms in \(H_3\). Hence the full density satisfies
\begin{align}
[H_1,h_3(x)]
=
i\,\partial_x h_3(x).    
\end{align}

Therefore,
\begin{align}
    [H_1,H_3]
=
\int dx\,[H_1,h_3(x)]
=
i\int dx\,\partial_x h_3(x).
\end{align}

Assuming periodic boundary conditions, or fields decaying sufficiently fast at infinity,
\begin{align}
\int dx\,\partial_x h_3(x)=0.    
\end{align}

Thus,
\begin{align}
[H_3,H_1]=0.   
\end{align}

\subsubsection{$[H_2,H_3]$}

We write
\begin{align}
H_2=A+\gamma B+\beta C,
\qquad
H_3=D+\gamma E+\beta G,    
\end{align}
where
\begin{align}
&A
=
-\sum_{m\geq 1}\frac{1}{m}
\int dx\,
\Psi_m^\dagger(x)\partial_x^2\Psi_m(x),\\
&B
=
\sum_{m\geq 1}m^3
\int dx\,
\Psi_m^\dagger(x)\Psi_m(x),\\
&C
=
\sum_{a,b\geq 1}
w_{a,b}
\int dx\,
\left(
\Psi_M^\dagger\Psi_a\Psi_b
+
\Psi_a^\dagger\Psi_b^\dagger\Psi_M
\right),
\end{align}
with
\begin{align}
    M=a+b,
\qquad
w_{a,b}=\sqrt{abM},
\qquad
a=m_1,\qquad
b=m_2.
\end{align}

The quadratic part of \(H_3\) are
\begin{align}
D
=
i\sum_{m\geq 1}\frac{1}{m^2}
\int dx\,
\Psi_m^\dagger\partial_x^3\Psi_m
\end{align}
and
\begin{align}
E
=
-3i\sum_{m\geq 1}m^2
\int dx\,
\Psi_m^\dagger\partial_x\Psi_m.    
\end{align}

The cubic part is
\begin{align}
    \begin{aligned}
G
=
i\sum_{a,b\geq 1}
w_{a,b}
\int dx\,
\Bigg[
&
\frac{1}{M}
(\partial_x\Psi_M^\dagger)\Psi_a\Psi_b
-\frac{1}{a}
\Psi_M^\dagger(\partial_x\Psi_a)\Psi_b
-\frac{1}{b}
\Psi_M^\dagger\Psi_a(\partial_x\Psi_b)
\\
&-
\frac{1}{M}
\Psi_a^\dagger\Psi_b^\dagger(\partial_x\Psi_M)
+\frac{1}{a}
(\partial_x\Psi_a^\dagger)\Psi_b^\dagger\Psi_M
+\frac{1}{b}
\Psi_a^\dagger(\partial_x\Psi_b^\dagger)\Psi_M
\Bigg].
\end{aligned}
\end{align}

Therefore
\begin{align}
    \begin{aligned}
[H_2,H_3]
=&
[A,D]
+\gamma\left([A,E]+[B,D]\right)
+\gamma^2[B,E]
\\
&+
\beta\left([A,G]+[C,D]\right)
+\beta\gamma\left([B,G]+[C,E]\right)
+\beta^2[C,G].
\end{aligned}
\end{align}

We show that each commutator (or sum of commutators) with coefficients $1, \gamma, \gamma^2, \beta, \beta \gamma, \beta^2$ vanishes.

First, the purely quadratic terms vanish. Indeed, \(A,D,B,E\) are all quadratic and diagonal in the mass index. Therefore their commutators are again quadratic. A direct integration by parts gives 
\begin{align}
&[A,D]=0,\\
&[A,E]+[B,D]=0,\\
&[B,E]=0.
\end{align}

We now consider the terms linear in \(\beta\).

\subsubsection*{The terms linear in \texorpdfstring{$\beta$}{beta}}

We shall prove that
\begin{align}
[A,G]+[C,D]=0.    
\end{align}

We split the $C$ and $G$ into a joining and splitting parts. The joining part of the cubic term \(C\) is
\begin{align}
    C^{(+)}
=
\sum_{a,b\geq 1}
w_{a,b}
\int dx\,
J_{a,b}(x),
\label{Cplus}
\end{align}
where
\begin{align}
    a=m_1,\qquad
b=m_2,\qquad
M=a+b,
\qquad
w_{a,b}=\sqrt{abM},
\end{align}
and $J_{a,b}(x)$ is \eqref{Jsplitting}.

The corresponding joining part of \(G\) is
\begin{align}
    G^{(+)}
=
i\sum_{a,b\geq 1}
w_{a,b}
\int dx\,
\mathcal R_{a,b}^{(+)}J_{a,b}(x),
\label{Gplus}
\end{align}
with
\begin{align}
    \mathcal R_{a,b}^{(+)}
=
\frac{1}{M}\partial_M
-\frac{1}{a}\partial_a
-\frac{1}{b}\partial_b.
\end{align}
Here the notation means
\begin{align}
&\partial_M J_{a,b}
=
(\partial_x\Psi_M^\dagger)\Psi_a\Psi_b,
&&\partial_a J_{a,b}
=
\Psi_M^\dagger(\partial_x\Psi_a)\Psi_b,
\qquad \partial_b J_{a,b}
=
\Psi_M^\dagger\Psi_a(\partial_x\Psi_b).
\end{align}

We first compute the action of \(A\) on the joining vertex. Using
\begin{align}
    [A,\Psi_m]
=
\frac{1}{m}\partial_x^2\Psi_m,
\qquad
[A,\Psi_m^\dagger]
=
-\frac{1}{m}\partial_x^2\Psi_m^\dagger,
\end{align}
we obtain
\begin{align}
    [A,J_{a,b}]
=
\mathcal E_{a,b}J_{a,b},
\end{align}
where
\begin{align}
    \mathcal E_{a,b}
=
-\frac{1}{M}\partial_M^2
+\frac{1}{a}\partial_a^2
+\frac{1}{b}\partial_b^2.
\end{align}

Since $A$ has no explicit dependence on $x$, the differential operator $\mathcal R_{a,b}^{(+)}$ may be
pulled through the commutator,
\begin{align}
    [A,G^{(+)}]
=
i\sum_{a,b\geq 1}
w_{a,b}
\int dx\,
\mathcal R_{a,b}^{(+)}\mathcal E_{a,b}J_{a,b}.
\end{align}

Now we compute the action of \(D\) on the same vertex. Using
\begin{align}
    [D,\Psi_m]
=
-\frac{i}{m^2}\partial_x^3\Psi_m,
\qquad
[D,\Psi_m^\dagger]
=
-\frac{i}{m^2}\partial_x^3\Psi_m^\dagger,
\end{align}
we get
\begin{align}
    [D,J_{a,b}]
=
-i\,\mathcal S_{a,b}J_{a,b},
\end{align}
where
\begin{align}
    \mathcal S_{a,b}
=
\frac{1}{M^2}\partial_M^3
+\frac{1}{a^2}\partial_a^3
+\frac{1}{b^2}\partial_b^3.
\end{align}
Therefore
\begin{align}
[C^{(+)},D]
=-[D,C^{(+)}]=
i\sum_{a,b\geq 1}
w_{a,b}
\int dx\,
\mathcal S_{a,b}J_{a,b}.
\end{align}

Hence
\begin{align}
    [A,G^{(+)}]+[C^{(+)},D]
=
i\sum_{a,b\geq 1}
w_{a,b}
\int dx\,
\left(
\mathcal R_{a,b}^{(+)}\mathcal E_{a,b}
+
\mathcal S_{a,b}
\right)J_{a,b}.
\end{align}

It remains to show that inside the integral
\begin{align}
    \mathcal R_{a,b}^{(+)}\mathcal E_{a,b}
+
\mathcal S_{a,b}
=0.
\end{align}

Inside an integrated vertex, total derivatives vanish, because fields are taken to be decaying sufficiently fast at the boundary. Therefore
\begin{align}
    \int dx\,
(\partial_M+\partial_a+\partial_b)J_{a,b}
=
\int dx\,\partial_x J_{a,b}
=0.
\end{align}
Thus, inside the integral, we may use
\begin{align}
    \partial_M\simeq -\partial_a-\partial_b,
    \label{partialMpartialab}
\end{align}
where \(\simeq\) means equality up to a total \(x\)-derivative. 

Let
\begin{align}
    u=\partial_a,
\qquad
v=\partial_b,
\qquad
\partial_M\simeq -(u+v).
\end{align}

Using \(M=a+b\), we find
\begin{align}
    \mathcal R_{a,b}^{(+)}
\simeq
-\frac{u+v}{M}
-\frac{u}{a}
-\frac{v}{b},
\qquad
\mathcal E_{a,b}
\simeq
-\frac{(u+v)^2}{M}
+\frac{u^2}{a}
+\frac{v^2}{b},
\qquad
\mathcal S_{a,b}
\simeq
-\frac{(u+v)^3}{M^2}
+\frac{u^3}{a^2}
+\frac{v^3}{b^2}
\end{align}

and a direct algebraic simplification gives
\begin{align}
    \begin{aligned}
&
\mathcal R_{a,b}^{(+)}\mathcal E_{a,b}
+
\mathcal S_{a,b}
\simeq\left(
-\frac{u+v}{M}
-\frac{u}{a}
-\frac{v}{b}
\right)
\left(
-\frac{(u+v)^2}{M}
+\frac{u^2}{a}
+\frac{v^2}{b}
\right)
\\
&\qquad
-\frac{(u+v)^3}{M^2}
+\frac{u^3}{a^2}
+\frac{v^3}{b^2}
=0.
\end{aligned}
\end{align}

Therefore
\begin{align}
    [A,G^{(+)}]+[C^{(+)},D]=0.
\end{align}

The splitting part is the Hermitian conjugate of the joining part, so the same cancellation holds
\begin{align}
    [A,G^{(-)}]+[C^{(-)},D]=0.
\end{align}

Combining the joining and splitting contributions, we obtain that the term linear in \(\beta\) vanishes
\begin{align}
\beta\left([A,G]+[C,D]\right)=0.
\end{align}

\subsubsection*{The term proportional to \texorpdfstring{$\beta\gamma$}{beta gamma}}

For the term proportional to \(\beta\gamma\), we want to prove that
\begin{align}
[B,G]+[C,E]=0.
\end{align}

First, let us compute the action of \(B\) on the joining vertex. Since
\begin{align}
    [B,\Psi_m]
=
-m^3\Psi_m,
\qquad
[B,\Psi_m^\dagger]
=
m^3\Psi_m^\dagger,
\end{align}
we get
\begin{align}
    [B,J_{a,b}]
=
\left(M^3-a^3-b^3\right)J_{a,b}=3abM J_{a,b}.
\end{align}

Since \(B\) does not contain \(x\)-derivatives, it commutes with the differential operator \(\mathcal R_{a,b}^{(+)}\). Hence
\begin{align}
    [B,G^{(+)}]
=
i\sum_{a,b\geq1}
w_{a,b}
\int dx\,
3abM\,\mathcal R_{a,b}^{(+)}J_{a,b}.
\end{align}

Now, we compute the action of \(E\) on the same joining vertex. Since
\begin{align}
    [E,\Psi_m]
=
3im^2\partial_x\Psi_m,
\qquad
[E,\Psi_m^\dagger]
=
3im^2\partial_x\Psi_m^\dagger,
\end{align}
we find
\begin{align}
    [E,J_{a,b}]
=
3i\,\mathcal L_{a,b}J_{a,b},
\end{align}
where
\begin{align}
    \mathcal L_{a,b}
=
M^2\partial_M
+a^2\partial_a
+b^2\partial_b.
\end{align}

Therefore
\begin{align}
    [C^{(+)},E]
=
-[E,C^{(+)}]
=
-3i\sum_{a,b\geq1}
w_{a,b}
\int dx\,
\mathcal L_{a,b}J_{a,b}.
\end{align}

Combining the two contributions, we obtain
\begin{align}
    [B,G^{(+)}]+[C^{(+)},E]
=
i\sum_{a,b\geq1}
w_{a,b}
\int dx\,
\left(
3abM\,\mathcal R_{a,b}^{(+)}
-
3\mathcal L_{a,b}
\right)
J_{a,b}.
\end{align}

It remains to show that inside the integral
\begin{align}
    abM\,\mathcal R_{a,b}^{(+)}
-
\mathcal L_{a,b}
\simeq 0.
\end{align}
By using \eqref{partialMpartialab}, we compute
\[
abM\,\mathcal R_{a,b}^{(+)}
=
abM
\left(
\frac{1}{M}\partial_M
-\frac{1}{a}\partial_a
-\frac{1}{b}\partial_b
\right)
=
ab\,\partial_M
-bM\,\partial_a
-aM\,\partial_b.
\]
Therefore
\begin{align}
    \begin{aligned}
        abM\,\mathcal R_{a,b}^{(+)}
-\mathcal L_{a,b}
=&
(ab-M^2)\partial_M
-(bM+a^2)\partial_a
-(aM+b^2)\partial_b.
\simeq\\
&-(ab-M^2)(\partial_a+\partial_b)
-(bM+a^2)\partial_a
-(aM+b^2)\partial_b
\\
&=
\left(
-ab+M^2-bM-a^2
\right)\partial_a
+
\left(
-ab+M^2-aM-b^2
\right)\partial_b.
    \end{aligned}
\end{align}

Now use \(M=a+b\). For the coefficient of \(\partial_a\),
\begin{align}
-ab+M^2-bM-a^2
=
-ab+M(M-b)-a^2
=
-ab+Ma-a^2=0.
\end{align}
Similarly, for the coefficient of \(\partial_b\),
\begin{align}
-ab+M^2-aM-b^2
=
-ab+M(M-a)-b^2
=
-ab+Mb-b^2
=
0.    
\end{align}

Therefore
\begin{align}
abM\,\mathcal R_{a,b}^{(+)}
-\mathcal L_{a,b}
\simeq 0    
\end{align}
and hence
\begin{align}
    [B,G^{(+)}]+[C^{(+)},E]=0.
\end{align}

The splitting part is the Hermitian conjugate of the joining part, so the same cancellation holds:
\begin{align}
[B,G^{(-)}]+[C^{(-)},E]=0.    
\end{align}

Combining joining and splitting contributions, we obtain that the term proportional to \(\beta\gamma\) vanishes:
\begin{align}
{[B,G]+[C,E]=0.}
\end{align}

\subsubsection*{The term proportional to \texorpdfstring{$\beta^2$}{beta squared}}
We want to show that
\begin{align}
    [C,G]=0.
\end{align}

We split the cubic terms into joining and splitting parts:
\begin{align}
    C=C^{(+)}+C^{(-)},
\qquad
G=G^{(+)}+G^{(-)}.
\end{align}
The joining part of \(C\) is \eqref{Cplus} and the splitting part is
\begin{align}
    C^{(-)}
=
\sum_{a,b\geq1}
w_{a,b}
\int dx\,
J_{a,b}^\dagger(x),
\end{align}
with $J_{a,b}^\dagger(x)$ given in \eqref{Jsplitting}.

Similarly, the joining part of \(G\) is given in \eqref{Gplus} and the splitting part is
\begin{align}
    G^{(-)}
=
i\sum_{c,d\geq1}
w_{c,d}
\int dy\,
\mathcal R_{c,d}^{(-)}J_{c,d}^\dagger(y),
\end{align}

where
\begin{align}
\begin{aligned}
\mathcal R_{c,d}^{(-)}
=&
-\frac{1}{c+d}\partial_{c+d}
+\frac{1}{c}\partial_c+\frac{1}{d}\partial_d.
\label{Rminus}
\end{aligned}
\end{align}

Thus
\begin{align}
[C,G]
=
[C^{(+)},G^{(+)}]
+
[C^{(-)},G^{(-)}]
+
[C^{(+)},G^{(-)}]
+
[C^{(-)},G^{(+)}].
\end{align}
We consider these four terms separately.

\paragraph{Joining--joining sector.}

We start from:
\begin{align}
    [C^{(+)},G^{(+)}]
=
i\sum_{a,b\geq1}
\sum_{c,d\geq1}
w_{a,b}w_{c,d}
\int dx\,dy\,
\left[
J_{a,b}(x),
\mathcal R_{c,d}^{(+)}J_{c,d}(y)
\right].
\end{align}

The only nonzero contractions are those in which an annihilation operator in one vertex meets the creation operator in the other vertex. We use the formula
\begin{align}
\begin{aligned}
[ABC,DFG]
={}&
AB[C,D]FG
+ABD[C,F]G
+ABDF[C,G]
\\
&+A[B,D]FGC
+AD[B,F]GC
+ADF[B,G]C
\\
&+[A,D]FGBC
+D[A,F]GBC
+DF[A,G]BC 
\end{aligned}
\label{commutatori33}
\end{align}
where
\begin{align}
    A=\Psi^\dagger_{a+b}(x),\qquad
B=\Psi_a(x),\qquad
C=\Psi_b(x),
\end{align}
\begin{align}
D=\Psi^\dagger_{c+d}(y),\qquad
F=\Psi_c(y),\qquad
G=\Psi_d(y).
\end{align}

We use the fact that
\begin{align}
\left[
J_{a,b}(x),
R^{(+)}_{c,d} J_{c,d}(y)
\right]
=
R^{(+)}_{c,d}
\left[
J_{a,b}(x),
J_{c,d}(y)
\right].
\end{align}
In \eqref{commutatori33}, 
\begin{align}
    [C,F]=[C,G]=[B,F]=[B,G]=[A,D]=0,
\end{align}
so the only non-zero terms are
\begin{align}
    [ABC,DFG]
=
AB[C,D]FG
+
A[B,D]FGC
+
D[A,F]GBC
+
DF[A,G]BC.
\end{align}

We explicitly compute the first term
\begin{align}
    \begin{aligned}
AB[C,D]FG
={}&
\Psi^\dagger_{a+b}(x)\Psi_a(x)
\left[
\Psi_b(x),\Psi^\dagger_{c+d}(y)
\right]
\Psi_c(y)\Psi_d(y)
\\
={}&
\delta_{b,c+d}\,\delta(x-y)\,
\Psi^\dagger_{a+b}(x)\Psi_a(x)
\Psi_c(y)\Psi_d(y)\\
={}&\delta(x-y)\,
\Psi^\dagger_{a+c+d}(x)
\Psi_a(x)\Psi_c(y)\Psi_d(y).
\end{aligned}
\end{align}
We relabel 
\begin{align}
    c\rightarrow b,\qquad
d\rightarrow c,\qquad
T=a+b+c.
\end{align}

So, we obtains a quartic joining vertex of the form
\begin{align}
Q_{a,b,c}^{(+)}(x)
=
\Psi_T^\dagger(x)\Psi_a(x)\Psi_b(x)\Psi_c(x),
\qquad
T=a+b+c.    
\end{align}

In this way of organising the calculation we may call this contribution the channel $(b,c)+a$, in reference to the combination which occurs between the labels after the re-indexing. The name of ``channel" does not bear any invariant meaning (due to the fact that the fields $\Psi$ can be commuted among themselves), and it is merely a way of streamlining the counting of the terms.

The weights are
\begin{align}
    \begin{aligned}
w_{a,b}\,w_{c,d}
\big|_{b=c+d}
&\to
w_{a,b+c}\,w_{b,c}
=
\sqrt{a(b+c)T}\,
\sqrt{b c(b+c)}
=
(b+c)\sqrt{abcT}.
\end{aligned}
\end{align}

Adding also the contribution of $R^{(+)}$ we obtain
\begin{align}
    AB[C,D]FG
\;\longrightarrow\;
(b+c)\sqrt{abcT}\,
R^{(a)}_{b,c}\,
Q^{(+)}_{a,b,c},
\end{align}
where 
\begin{align}
    R^{(a)}_{b,c}
=
\frac{1}{b+c}
\left(
\partial_T+\partial_a
\right)
-\frac{1}{b}\partial_b
-\frac{1}{c}\partial_c.
\end{align}
The reason of this $R_{b,c}^{(a)}$ is that the differential operator associated with the \((b,c)\) vertex before
the contraction is
\begin{align}
    R^{(+)}_{b,c}
=
\frac{1}{b+c}\partial_{b+c}
-\frac{1}{b}\partial_b
-\frac{1}{c}\partial_c.
\end{align}

The last two derivatives act on fields which are not contracted, and
therefore remain unchanged:
\begin{align}
    -\frac{1}{b}\partial_b-\frac{1}{c}\partial_c.
\end{align}

On the other hand, \(\partial_{b+c}\) acts on the field
\(\Psi^\dagger_{b+c}\) which is contracted. Hence it produces a
derivative of the Dirac delta,
\begin{align}
    \left[
\Psi_{b+c}(x),
\partial_y\Psi^\dagger_{b+c}(y)
\right]
=
\partial_y\delta(x-y).
\end{align}

Under the integral, integration by parts gives
\begin{align}
    \begin{aligned}
\int dx\,dy\,
\Psi^\dagger_T(x)\Psi_a(x)
\Psi_b(y)\Psi_c(y)\,
\partial_y\delta(x-y)
&=
-\int dx\,
\Psi^\dagger_T\Psi_a\,
\partial_x(\Psi_b\Psi_c)\simeq
\int dx\,
\partial_x(\Psi^\dagger_T\Psi_a)\,
\Psi_b\Psi_c ,
\end{aligned}
\end{align}
where \(\simeq\) denotes equality modulo a total derivative.

Therefore,
\begin{align}
    \partial_{b+c}
\;\longrightarrow\;
\partial_T+\partial_a,
\end{align}

and the effective differential operator in the channel
\((b,c)+a\) is
\begin{align}
    R^{(a)}_{b,c}
=
\frac{1}{b+c}
\left(
\partial_T+\partial_a
\right)
-\frac{1}{b}\partial_b
-\frac{1}{c}\partial_c .
\end{align}

We continue to use the convention whereby the derivative with an index singles out the variable of the corresponding field. In this way the differential operators $R$ acquire a specification, which we symbolise with the addition of book-keeping indices. We recall that in expressions obtained after a contraction, such as
$\mathcal R^{(\pm)}_{a,b}[\cdots]$, the differential operator is
understood as inherited from the corresponding cubic vertex before
the contraction is performed. In particular, if a derivative acts
on the contracted field, it acts on the Dirac delta produced by the
contraction.

In the same way, we can calculate the remaining three terms.

Let us now consider the second non-vanishing contribution,
\begin{align}
\begin{aligned}
A[B,D]FGC
={}&
\Psi^\dagger_{a+b}(x)
\left[
\Psi_a(x),\Psi^\dagger_{c+d}(y)
\right]
\Psi_c(y)\Psi_d(y)\Psi_b(x)
\\
={}&
\delta_{a,c+d}\,\delta(x-y)\,
\Psi^\dagger_{a+b}(x)
\Psi_c(y)\Psi_d(y)\Psi_b(x)
\\
={}&
\delta(x-y)\,
\Psi^\dagger_{b+c+d}(x)
\Psi_c(y)\Psi_d(y)\Psi_b(x).
\end{aligned}
\end{align}
In this case the Kronecker delta imposes
\begin{align}
a=c+d.
\end{align}
We relabel the remaining independent indices as
\begin{align}
b\rightarrow a,\qquad
c\rightarrow b,\qquad
d\rightarrow c,
\qquad
T=a+b+c.
\end{align}
Therefore, after using the Dirac delta and reordering the $\Psi$ fields, we obtain
again the quartic vertex
\begin{align}
Q^{(+)}_{a,b,c}(x)
=
\Psi_T^\dagger(x)\Psi_a(x)\Psi_b(x)\Psi_c(x).
\end{align}
This contribution corresponds again to the channel \((b,c)+a\).

The product of weights becomes
\begin{align}
w_{a,b}\,w_{c,d}
\big|_{a=c+d}\longrightarrow
w_{b+c,a}\,w_{b,c}=\sqrt{(b+c)aT}\,
\sqrt{bc(b+c)}=(b+c)\sqrt{abcT}.
\end{align}
Since the contracted field is again the creation operator of the
\((b,c)\) vertex, the corresponding effective differential operator
is the same as for the previous contribution,
\begin{align}
R^{(a)}_{b,c}
=
\frac{1}{b+c}
\left(
\partial_T+\partial_a
\right)
-\frac{1}{b}\partial_b
-\frac{1}{c}\partial_c.
\end{align}
Hence,
\begin{align}
A[B,D]FGC
\;\longrightarrow\;
(b+c)\sqrt{abcT}\,
R^{(a)}_{b,c}\,
Q^{(+)}_{a,b,c}.
\end{align}

We now consider the third non-vanishing contribution,
\begin{align}
\begin{aligned}
D[A,F]GBC
={}&
\Psi^\dagger_{c+d}(y)
\left[
\Psi^\dagger_{a+b}(x),\Psi_c(y)
\right]
\Psi_d(y)\Psi_a(x)\Psi_b(x)
\\
={}&
-\delta_{a+b,c}\,\delta(x-y)\,
\Psi^\dagger_{c+d}(y)
\Psi_d(y)\Psi_a(x)\Psi_b(x)
\\
={}&
-\delta(x-y)\,
\Psi^\dagger_{a+b+d}(y)
\Psi_d(y)\Psi_a(x)\Psi_b(x).
\end{aligned}
\end{align}
Here the Kronecker delta imposes
\begin{align}
c=a+b.
\end{align}
We relabel
\begin{align}
d\rightarrow c,
\qquad
T=a+b+c.
\end{align}
After using the Dirac delta, the remaining fields
again form
\begin{align}
Q^{(+)}_{a,b,c}(x)
=
\Psi_T^\dagger(x)\Psi_a(x)\Psi_b(x)\Psi_c(x).
\end{align}
This contribution belongs to the channel \((a,b)+c\).

The product of weights is
\begin{align}
w_{a,b}\,w_{c,d}
\big|_{c=a+b}\longrightarrow
w_{a,b}\,w_{a+b,c}
=
\sqrt{ab(a+b)}\,
\sqrt{(a+b)cT}
=
(a+b)\sqrt{abcT}.
\end{align}

Let us now determine the effective differential operator.
Before the contraction, the second vertex is of the form
\begin{align}
J_{a+b,c}
=
\Psi^\dagger_T\Psi_{a+b}\Psi_c,
\end{align}
and its differential operator is
\begin{align}
R^{(+)}_{a+b,c}
=
\frac{1}{T}\partial_T
-\frac{1}{a+b}\partial_{a+b}
-\frac{1}{c}\partial_c.
\end{align}
The derivatives \(\partial_T\) and \(\partial_c\) act on fields which
are not contracted, and therefore remain unchanged. On the other hand,
\(\partial_{a+b}\) acts on the contracted field \(\Psi_{a+b}\).
After the contraction, the corresponding derivative is transferred to
the two remaining fields \(\Psi_a\Psi_b\). Thus, modulo a total
derivative,
\begin{align}
\partial_{a+b}
\;\longrightarrow\;
\partial_a+\partial_b.
\end{align}
Therefore the effective differential operator is
\begin{align}
R^{(ab)}_{a+b,c}
=
\frac{1}{T}\partial_T
-\frac{1}{a+b}
\left(
\partial_a+\partial_b
\right)
-\frac{1}{c}\partial_c.
\end{align}

Hence,
\begin{align}
D[A,F]GBC
\;\longrightarrow\;
-(a+b)\sqrt{abcT}\,
R^{(ab)}_{a+b,c}\,
Q^{(+)}_{a,b,c}.
\end{align}

Finally, consider the fourth non-vanishing contribution,
\begin{align}
\begin{aligned}
DF[A,G]BC
={}&
\Psi^\dagger_{c+d}(y)\Psi_c(y)
\left[
\Psi^\dagger_{a+b}(x),\Psi_d(y)
\right]
\Psi_a(x)\Psi_b(x)
\\
={}&
-\delta_{a+b,d}\,\delta(x-y)\,
\Psi^\dagger_{c+d}(y)\Psi_c(y)
\Psi_a(x)\Psi_b(x)
\\
={}&
-\delta(x-y)\,
\Psi^\dagger_{a+b+c}(y)\Psi_c(y)
\Psi_a(x)\Psi_b(x).
\end{aligned}
\end{align}
The Kronecker delta now imposes
\begin{align}
d=a+b.
\end{align}
We label
\begin{align}
T=a+b+c.
\end{align}
The remaining fields form again
\begin{align}
Q^{(+)}_{a,b,c}(x)
=
\Psi_T^\dagger(x)\Psi_a(x)\Psi_b(x)\Psi_c(x),
\end{align}
and this contribution also belongs to the channel \((a,b)+c\).

The corresponding product of weights is
\begin{align}
w_{a,b}\,w_{c,d}
\big|_{d=a+b}\longrightarrow
w_{a,b}\,w_{c,a+b}
=
\sqrt{ab(a+b)}\,
\sqrt{c(a+b)T}
=
(a+b)\sqrt{abcT}.
\end{align}
Since the contracted field has again mass \(a+b\), the effective
differential operator is
\begin{align}
R^{(ab)}_{a+b,c}
=
\frac{1}{T}\partial_T
-\frac{1}{a+b}
\left(
\partial_a+\partial_b
\right)
-\frac{1}{c}\partial_c.
\end{align}
Therefore,
\begin{align}
DF[A,G]BC
\;\longrightarrow\;
-(a+b)\sqrt{abcT}\,
R^{(ab)}_{a+b,c}\,
Q^{(+)}_{a,b,c}.
\end{align}

To summarize, we obtain that the four non-vanishing contributions are
\begin{align}
AB[C,D]FG
&\longrightarrow
+(b+c)\sqrt{abcT}\,
R^{(a)}_{b,c}\,
Q^{(+)}_{a,b,c},
\\[2mm]
A[B,D]FGC
&\longrightarrow
+(b+c)\sqrt{abcT}\,
R^{(a)}_{b,c}\,
Q^{(+)}_{a,b,c},
\\[2mm]
D[A,F]GBC
&\longrightarrow
-(a+b)\sqrt{abcT}\,
R^{(ab)}_{a+b,c}\,
Q^{(+)}_{a,b,c},
\\[2mm]
DF[A,G]BC
&\longrightarrow
-(a+b)\sqrt{abcT}\,
R^{(ab)}_{a+b,c}\,
Q^{(+)}_{a,b,c},
\end{align}
where
\begin{align}
Q^{(+)}_{a,b,c}
=
\Psi_T^\dagger\Psi_a\Psi_b\Psi_c,
\qquad
T=a+b+c,
\end{align}
and
\begin{align}
R^{(a)}_{b,c}
&=
\frac{1}{b+c}
\left(
\partial_T+\partial_a
\right)
-\frac{1}{b}\partial_b
-\frac{1}{c}\partial_c,
\\
R^{(ab)}_{a+b,c}
&=
\frac{1}{T}\partial_T
-\frac{1}{a+b}
\left(
\partial_a+\partial_b
\right)
-\frac{1}{c}\partial_c.
\end{align}

At this stage, the four contractions give
\begin{align}
\mathcal K_{a,b,c}
=
2(b+c)R^{(a)}_{b,c}
-
2(a+b)R^{(ab)}_{a+b,c}.
\end{align}
Notice, however, that this operator does not need to be a total
derivative by itself. Indeed, it appears inside the completely
symmetric sum
\begin{align}
\sum_{a,b,c\geq 1}
\sqrt{abcT}\int dx\,
\mathcal K_{a,b,c}\,
Q^{(+)}_{a,b,c},
\qquad T=a+b+c,
\end{align}
and both the weight and the quartic vertex are invariant under
permutations of the dummy indices \(a,b,c\).

We can therefore perform the cyclic relabellings
\begin{align}
(a,b,c)\rightarrow(b,c,a),
\qquad
(a,b,c)\rightarrow(c,a,b),
\end{align}
which give
\begin{align}
\mathcal K_{b,c,a}
&=
2(a+c)R^{(b)}_{a,c}
-
2(b+c)R^{(bc)}_{b+c,a},
\\
\mathcal K_{c,a,b}
&=
2(a+b)R^{(c)}_{a,b}
-
2(a+c)R^{(ac)}_{a+c,b}.
\end{align}

Since the three expressions give the same contribution after summation
over the dummy indices, we may replace \(\mathcal K_{a,b,c}\) by its
cyclic average,
\begin{align}
\frac{1}{3}
\left(
\mathcal K_{a,b,c}
+
\mathcal K_{b,c,a}
+
\mathcal K_{c,a,b}
\right)
=
\frac{2}{3}\mathcal A^{(+)}_{a,b,c},
\end{align}
where
\begin{align}
\mathcal A^{(+)}_{a,b,c}
=
(a+b)
\left(
R^{(c)}_{a,b}
-
R^{(ab)}_{a+b,c}
\right)
+
(a+c)
\left(
R^{(b)}_{a,c}
-
R^{(ac)}_{a+c,b}
\right)
+
(b+c)
\left(
R^{(a)}_{b,c}
-
R^{(bc)}_{b+c,a}
\right).
\end{align}

Substituting the explicit expressions of the differential operators,
we find
\begin{align}
(a+b)
\left(
R^{(c)}_{a,b}
-
R^{(ab)}_{a+b,c}
\right)
&=
\frac{c}{T}\partial_T
-\frac{b}{a}\partial_a
-\frac{a}{b}\partial_b
+\frac{T}{c}\partial_c,
\\
(a+c)
\left(
R^{(b)}_{a,c}
-
R^{(ac)}_{a+c,b}
\right)
&=
\frac{b}{T}\partial_T
-\frac{c}{a}\partial_a
+\frac{T}{b}\partial_b
-\frac{a}{c}\partial_c,
\\
(b+c)
\left(
R^{(a)}_{b,c}
-
R^{(bc)}_{b+c,a}
\right)
&=
\frac{a}{T}\partial_T
+\frac{T}{a}\partial_a
-\frac{c}{b}\partial_b
-\frac{b}{c}\partial_c.
\end{align}
Therefore, using \(T=a+b+c\),
\begin{align}
\mathcal A^{(+)}_{a,b,c}
=
\partial_T+\partial_a+\partial_b+\partial_c.
\end{align}
Hence
\begin{align}
\mathcal A^{(+)}_{a,b,c}
Q^{(+)}_{a,b,c}
=
\partial_x Q^{(+)}_{a,b,c},
\end{align}
and the joining--joining contribution vanishes after integration over
\(x\), assuming the usual boundary conditions.

\paragraph{Splitting--splitting sector.}

The splitting--splitting sector is the Hermitian conjugate of the joining--joining sector. Therefore the same argument gives
\begin{align}
[C^{(-)},G^{(-)}]=0.    
\end{align}

\paragraph{Mixed sectors.}
It remains to consider
\begin{equation}
    [C^{(+)},G^{(-)}]+[C^{(-)},G^{(+)}].
\end{equation}
We start from independent summation indices and independent positions,
\begin{equation}
[C^{(+)},G^{(-)}]
=
i\sum_{a,b\geq1}\sum_{c,d\geq1}
w_{a,b}w_{c,d}
\int dx\,dy\,
\left[
J_{a,b}(x),
\mathcal R^{(-)}_{c,d}J^\dagger_{c,d}(y)
\right],
\end{equation}
where $J_{a,b}(x)$ is given in \eqref{Jsplitting} and $R^{(-)}$ in \eqref{Rminus}.

\begin{equation}
J_{a,b}(x)
=
\Psi^\dagger_{a+b}(x)\Psi_a(x)\Psi_b(x),
\end{equation}
and
\begin{align}
\mathcal R^{(-)}_{c,d}J^\dagger_{c,d}(y)
=
-\frac{1}{c+d}
\Psi^\dagger_c(y)\Psi^\dagger_d(y)
\partial_y\Psi_{c+d}(y)
+
\frac{1}{c}
\left(\partial_y\Psi^\dagger_c(y)\right)
\Psi^\dagger_d(y)\Psi_{c+d}(y)
+
\frac{1}{d}
\Psi^\dagger_c(y)
\left(\partial_y\Psi^\dagger_d(y)\right)
\Psi_{c+d}(y).
\end{align}

As in the joining--joining sector, we first consider the commutator
of the undifferentiated cubic vertices. We set
\begin{equation}
A=\Psi^\dagger_{a+b}(x),
\qquad
B=\Psi_a(x),
\qquad
C=\Psi_b(x),
\end{equation}
and
\begin{equation}
D=\Psi^\dagger_c(y),
\qquad
F=\Psi^\dagger_d(y),
\qquad
G=\Psi_{c+d}(y).
\end{equation}
Using
\begin{align}
[ABC,DFG]
={}&
AB[C,D]FG+ABD[C,F]G+ABDF[C,G]
\nonumber\\
&+
A[B,D]FGC+AD[B,F]GC+ADF[B,G]C
\nonumber\\
&+
[A,D]FGBC+D[A,F]GBC+DF[A,G]BC,
\label{manycommutators}
\end{align}
we now have
\begin{equation}
[C,G]=[B,G]=[A,D]=[A,F]=0,
\end{equation}
whereas
\begin{align}
[C,D]&=\delta_{b,c}\delta(x-y),
&
[C,F]&=\delta_{b,d}\delta(x-y),
\nonumber\\
[B,D]&=\delta_{a,c}\delta(x-y),
&
[B,F]&=\delta_{a,d}\delta(x-y),
\nonumber\\
[A,G]&=-\delta_{a+b,c+d}\delta(x-y).
\end{align}
Therefore, we call the five non-vanishing terms $T_1,T_2,T_3,T_4,T_5$
\begin{align}
[ABC,DFG]
={}&
\underbrace{AB[C,D]FG}_{T_1}
+
\underbrace{ABD[C,F]G}_{T_2}
+
\underbrace{A[B,D]FGC}_{T_3}
\nonumber\\
&+
\underbrace{AD[B,F]GC}_{T_4}
+
\underbrace{DF[A,G]BC}_{T_5}.
\label{eq:mixed-five-contractions}
\end{align}

We now consider the opposite mixed commutator,
\begin{equation}
[C^{(-)},G^{(+)}]
=
i\sum_{a,b\geq1}\sum_{c,d\geq1}
w_{a,b}w_{c,d}
\int dx\,dy\,
\left[
J^\dagger_{a,b}(x),
\mathcal R^{(+)}_{c,d}J_{c,d}(y)
\right],
\end{equation}
where
\begin{equation}
J^\dagger_{a,b}(x)
=
\Psi^\dagger_a(x)\Psi^\dagger_b(x)\Psi_{a+b}(x),
\end{equation}
and
\begin{align}
\mathcal R^{(+)}_{c,d}J_{c,d}(y)
=&
\frac{1}{c+d}
\left(\partial_y\Psi^\dagger_{c+d}(y)\right)
\Psi_c(y)\Psi_d(y)
-
\frac{1}{c}
\Psi^\dagger_{c+d}(y)
\left(\partial_y\Psi_c(y)\right)
\Psi_d(y)
\nonumber \\&-
\frac{1}{d}
\Psi^\dagger_{c+d}(y)
\Psi_c(y)
\left(\partial_y\Psi_d(y)\right).
\end{align}

For the undifferentiated cubic vertices, we now set
\begin{equation}
A=\Psi^\dagger_a(x),
\qquad
B=\Psi^\dagger_b(x),
\qquad
C=\Psi_{a+b}(x),
\end{equation}
and
\begin{equation}
D=\Psi^\dagger_{c+d}(y),
\qquad
F=\Psi_c(y),
\qquad
G=\Psi_d(y).
\end{equation}
Using again
\begin{align}
[ABC,DFG]
={}&
AB[C,D]FG+ABD[C,F]G+ABDF[C,G]
\nonumber\\
&+
A[B,D]FGC+AD[B,F]GC+ADF[B,G]C
\nonumber\\
&+
[A,D]FGBC+D[A,F]GBC+DF[A,G]BC,
\end{align}
we have
\begin{equation}
[C,F]=[C,G]=[B,D]=[A,D]=0,
\end{equation}
whereas
\begin{align}
[C,D]
&=
\delta_{a+b,c+d}\delta(x-y),
\nonumber\\
[B,F]
&=
-\delta_{b,c}\delta(x-y),
&
[B,G]
&=
-\delta_{b,d}\delta(x-y),
\nonumber\\
[A,F]
&=
-\delta_{a,c}\delta(x-y),
&
[A,G]
&=
-\delta_{a,d}\delta(x-y).
\end{align}
Therefore, also in this case there are five non-vanishing
single-contraction terms, which we denote by
$\bar T_1,\bar T_2,\bar T_3,\bar T_4,\bar T_5$:
\begin{align}
[ABC,DFG]
={}&
\underbrace{AB[C,D]FG}_{\bar T_1}
+
\underbrace{AD[B,F]GC}_{\bar T_2}
+
\underbrace{ADF[B,G]C}_{\bar T_3}
\nonumber\\
&+
\underbrace{D[A,F]GBC}_{\bar T_4}
+
\underbrace{DF[A,G]BC}_{\bar T_5}.
\label{eq:mixed-five-contractions-opposite}
\end{align}

We consider separately three blocks
\begin{align}
   & B_1=T_1+T_2+\bar{T_4}+\bar{T}_5\\
   & B_2=T_3+T_4+\bar{T_2}+\bar{T}_3\\
   & B_3=T_5+\bar{T}_1
\end{align}

Let us start from $B_1$.

For the first term,
\begin{equation}
T_1=AB[C,D]FG,
\end{equation}
we use
\begin{equation}
[C,D]
=
[\Psi_b(x),\Psi^\dagger_c(y)]
=
\delta_{b,c}\delta(x-y).
\end{equation}
The Kronecker delta sets $c=b$, while the Dirac delta allows us to
perform the $y$-integration. Therefore the corresponding contribution
to $[C^{(+)},G^{(-)}]$ is
\begin{align}
T_1
=
i\sum_{a,b,d\geq1}
w_{a,b}w_{b,d}
\int dx\,
\mathcal R^{(-)}_{b,d}
\left[
\Psi^\dagger_{a+b}(x)\Psi_a(x)
\Psi^\dagger_d(x)\Psi_{b+d}(x)
\right].
\label{eq:mixed-T1}
\end{align}

For the second term,
\begin{equation}
T_2=ABD[C,F]G,
\end{equation}
we instead use
\begin{equation}
[C,F]
=
[\Psi_b(x),\Psi^\dagger_d(y)]
=
\delta_{b,d}\delta(x-y).
\end{equation}
Now the Kronecker delta sets $d=b$, and we obtain
\begin{align}
T_2
=
i\sum_{a,b,c\geq1}
w_{a,b}w_{c,b}
\int dx\,
\mathcal R^{(-)}_{c,b}
\left[
\Psi^\dagger_{a+b}(x)\Psi_a(x)
\Psi^\dagger_c(x)\Psi_{c+b}(x)
\right].
\label{eq:mixed-T2-before-relabel}
\end{align}

The summation index $c$ in the last expression is a dummy index.
We may therefore relabel it as
\begin{equation}
c\longrightarrow d.
\end{equation}
This gives
\begin{align}
T_2
=
i\sum_{a,b,d\geq1}
w_{a,b}w_{d,b}
\int dx\,
\mathcal R^{(-)}_{d,b}
\left[
\Psi^\dagger_{a+b}(x)\Psi_a(x)
\Psi^\dagger_d(x)\Psi_{d+b}(x)
\right].
\end{align}

We now use the symmetry
\begin{equation}
w_{d,b}=w_{b,d},
\end{equation}
and
\begin{equation}
\mathcal R^{(-)}_{d,b}J^\dagger_{d,b}
=
\mathcal R^{(-)}_{b,d}J^\dagger_{b,d}.
\label{eq:Rminus-symmetry}
\end{equation}

Consequently,
\begin{equation}
T_2=T_1,
\end{equation}
and hence
\begin{align}
T_1+T_2
=
2i\sum_{a,b,d\geq1}
w_{a,b}w_{b,d}
\int dx\,
\mathcal R^{(-)}_{b,d}
\left[
\Psi^\dagger_{a+b}\Psi_a
\Psi^\dagger_d\Psi_{b+d}
\right].
\label{eq:mixed-T1-T2}
\end{align}

We now consider $\bar T_4$ and $\bar T_5$.
For
\begin{equation}
\bar T_4=D[A,F]GBC,
\end{equation}
we have
\begin{equation}
[A,F]
=
[\Psi^\dagger_a(x),\Psi_c(y)]
=
-\delta_{a,c}\delta(x-y),
\end{equation}
so that $c=a$ and
\begin{align}
\bar T_4
=
-i\sum_{a,b,d\geq1}
w_{a,b}w_{a,d}
\int dx\,
\mathcal R^{(+)}_{a,d}
\left[
\Psi^\dagger_{a+d}\Psi_d
\Psi^\dagger_b\Psi_{a+b}
\right].
\label{eq:mixed-barT4}
\end{align}

Similarly, for
\begin{equation}
\bar T_5=DF[A,G]BC,
\end{equation}
we use
\begin{equation}
[A,G]
=
[\Psi^\dagger_a(x),\Psi_d(y)]
=
-\delta_{a,d}\delta(x-y),
\end{equation}
and therefore
\begin{align}
\bar T_5
=
-i\sum_{a,b,c\geq1}
w_{a,b}w_{c,a}
\int dx\,
\mathcal R^{(+)}_{c,a}
\left[
\Psi^\dagger_{c+a}\Psi_c
\Psi^\dagger_b\Psi_{a+b}
\right].
\end{align}

Relabeling the dummy index $c\to d$ and using the symmetry, we can readjust this expression,

\begin{equation}
w_{d,a}=w_{a,d},
\qquad
\mathcal R^{(+)}_{d,a}J_{d,a}
=
\mathcal R^{(+)}_{a,d}J_{a,d},
\end{equation}
we find
\begin{equation}
\bar T_5=\bar T_4.
\end{equation}
Hence
\begin{align}
\bar T_4+\bar T_5
=
-2i\sum_{a,b,d\geq1}
w_{a,b}w_{a,d}
\int dx\,
\mathcal R^{(+)}_{a,d}
\left[
\Psi^\dagger_{a+d}\Psi_d
\Psi^\dagger_b\Psi_{a+b}
\right].
\label{eq:mixed-barT4-barT5}
\end{align}

In order to combine this expression with \eqref{eq:mixed-T1-T2},
we perform in \eqref{eq:mixed-barT4-barT5} the simultaneous
relabeling of dummy indices
\begin{equation}
d\to a,
\qquad
b\to d,
\qquad
a\to b.
\end{equation}
\begin{align}
\bar T_4+\bar T_5
=
-2i\sum_{a,b,d\geq1}
w_{a,b}w_{b,d}
\int dx\,
\mathcal R^{(+)}_{b,a}
\left[
\Psi^\dagger_{a+b}\Psi_a
\Psi^\dagger_d\Psi_{b+d}
\right].
\end{align}

Introducing 
\begin{equation}
\widetilde Q_{a,b+d}^{a+b,d}(x)
=
\Psi^\dagger_{a+b}(x)\Psi_a(x)
\Psi^\dagger_d(x)\Psi_{b+d}(x),
\end{equation}
we finally obtain
\begin{align}
T_1+T_2+\bar T_4+\bar T_5
=
2i\sum_{a,b,d\geq1}
w_{a,b}w_{b,d}
\int dx\,
\left(
\mathcal R^{(-)}_{b,d}
-
\mathcal R^{(+)}_{b,a}
\right)
\widetilde Q_{a,b+d}^{a+b,d}(x).
\label{eq:first-mixed-block}
\end{align}

Let us now make explicit the action of the differential operators
in \eqref{eq:first-mixed-block}. We introduce, for convenience,
\begin{equation}
U(x)=\Psi^\dagger_{a+b}(x)\Psi_a(x),
\qquad
V(x)=\Psi^\dagger_d(x)\Psi_{b+d}(x),
\end{equation}
so that
\begin{equation}
\widetilde Q_{a,b+d}^{a+b,d}(x)=U(x)V(x).
\end{equation}

We first consider $\mathcal R^{(-)}_{b,d}$. The derivatives acting
on the uncontracted fields remain unchanged. On the other hand,
the term proportional to $1/b$ contains a derivative acting on the
contracted field $\Psi_b^\dagger$. Since
\begin{equation}
[\Psi_b(x),\partial_y\Psi_b^\dagger(y)]
=
\partial_y\delta(x-y),
\end{equation}
we find
\begin{align}
\int dx\,dy\,
U(x)V(y)\partial_y\delta(x-y)
&=
-\int dx\,U(x)\partial_xV(x)
\simeq
\int dx\,(\partial_xU(x))V(x),
\end{align}
where $\simeq$ denotes equality modulo a total $x$-derivative.
Thus
\begin{equation}
\partial_b^{\rm contr.}
\longrightarrow
\partial_{a+b}+\partial_a,
\end{equation}
and therefore 
\begin{align}
\mathcal R^{(-)}_{b,d}
\rightsquigarrow
-\frac{1}{b+d}\partial_{b+d}
+
\frac{1}{b}
\left(\partial_{a+b}+\partial_a\right)
+
\frac{1}{d}\partial_d.
\label{eq:Rminus-B1-effective}
\end{align}

We now consider the contribution coming from
$-\mathcal R^{(+)}_{b,a}$. The derivative proportional to $1/b$
again acts on the contracted field. In this case
\begin{equation}
[\Psi_b^\dagger(x),\partial_y\Psi_b(y)]
=
-\partial_y\delta(x-y).
\end{equation}
Combining this sign with the coefficient $-1/b$ in
$\mathcal R^{(+)}_{b,a}$ gives a positive derivative of the Dirac
delta. After the relabelling used above, the corresponding integral
can be written as
\begin{align}
\frac{1}{b}
\int dx\,dy\,
U(y)V(x)\partial_y\delta(x-y)
&=
-\frac{1}{b}
\int dx\,(\partial_xU)V
\simeq
\frac{1}{b}
\int dx\,U(\partial_xV).
\end{align}
Hence the effective contribution of $-\mathcal R^{(+)}_{b,a}$ is
\begin{align}
-\mathcal R^{(+)}_{b,a}
\rightsquigarrow
-\frac{1}{a+b}\partial_{a+b}
+
\frac{1}{b}
\left(\partial_d+\partial_{b+d}\right)
+
\frac{1}{a}\partial_a.
\label{eq:Rplus-B1-effective}
\end{align}

Combining \eqref{eq:Rminus-B1-effective} and
\eqref{eq:Rplus-B1-effective}, we obtain 

\begin{align}
B_1
={}&
2i\sum_{a,b,d\geq1}
w_{a,b}w_{b,d}
\int dx\,\Bigg[
-\frac{1}{b+d}
\Psi^\dagger_{a+b}\Psi_a\Psi^\dagger_d
(\partial_x\Psi_{b+d})
\nonumber\\
&\hspace{25mm}
+\frac{1}{b}
\partial_x\!\left(
\Psi^\dagger_{a+b}\Psi_a
\right)
\Psi^\dagger_d\Psi_{b+d}
+\frac{1}{d}
\Psi^\dagger_{a+b}\Psi_a
(\partial_x\Psi^\dagger_d)\Psi_{b+d}
-\frac{1}{a+b}
(\partial_x\Psi^\dagger_{a+b})
\Psi_a\Psi^\dagger_d\Psi_{b+d}
\nonumber\\
&\hspace{25mm}
+\frac{1}{b}
\Psi^\dagger_{a+b}\Psi_a
\partial_x\!\left(
\Psi^\dagger_d\Psi_{b+d}
\right)
+\frac{1}{a}
\Psi^\dagger_{a+b}
(\partial_x\Psi_a)
\Psi^\dagger_d\Psi_{b+d}
\Bigg]\simeq \nonumber\\
{}&
2i\sum_{a,b,d\geq1}
w_{a,b}w_{b,d}
\int dx\,\Bigg[
-\frac{1}{b+d}
\Psi^\dagger_{a+b}\Psi_a\Psi^\dagger_d
(\partial_x\Psi_{b+d})
\nonumber\\
&\hspace{25mm}
+\frac{1}{d}
\Psi^\dagger_{a+b}\Psi_a
(\partial_x\Psi^\dagger_d)\Psi_{b+d}
-\frac{1}{a+b}
(\partial_x\Psi^\dagger_{a+b})
\Psi_a\Psi^\dagger_d\Psi_{b+d}
+\frac{1}{a}
\Psi^\dagger_{a+b}
(\partial_x\Psi_a)
\Psi^\dagger_d\Psi_{b+d}
\Bigg].
\label{eq:B1-expanded}
\end{align}

We omitted the terms with $1/b$ since they are proportional to a total derivative.

We now consider $B_2$
\begin{equation}
T_3=A[B,D]FGC,
\end{equation}
we use
\begin{equation}
[B,D]
=
[\Psi_a(x),\Psi_c^\dagger(y)]
=
\delta_{a,c}\delta(x-y),
\end{equation}
so that $c=a$. Hence
\begin{align}
T_3
=
i\sum_{a,b,d\geq1}
w_{a,b}w_{a,d}
\int dx\,
\mathcal R^{(-)}_{a,d}
\left[
\Psi^\dagger_{a+b}
\Psi^\dagger_d
\Psi_{a+d}
\Psi_b
\right].
\end{align}
Similarly,
\begin{equation}
T_4=AD[B,F]GC,
\qquad
[B,F]=\delta_{a,d}\delta(x-y).
\end{equation}

After setting $d=a$, relabelling $c\to d$, and using the
symmetries
\begin{equation}
w_{d,a}=w_{a,d},
\qquad
\mathcal R^{(-)}_{d,a}J^\dagger_{d,a}
=
\mathcal R^{(-)}_{a,d}J^\dagger_{a,d},
\end{equation}
we obtain $T_4=T_3$. Therefore
\begin{align}
T_3+T_4
=
2i\sum_{a,b,d\geq1}
w_{a,b}w_{a,d}
\int dx\,
\mathcal R^{(-)}_{a,d}
\left[
\Psi^\dagger_{a+b}
\Psi^\dagger_d
\Psi_{a+d}
\Psi_b
\right].
\label{eq:mixed-T3-T4}
\end{align}

For the opposite mixed commutator,
\begin{equation}
\bar T_2=AD[B,F]GC,
\qquad
[B,F]=-\delta_{b,c}\delta(x-y),
\end{equation}
and hence
\begin{align}
\bar T_2
=
-i\sum_{a,b,d\geq1}
w_{a,b}w_{b,d}
\int dx\,
\mathcal R^{(+)}_{b,d}
\left[
\Psi^\dagger_a
\Psi^\dagger_{b+d}
\Psi_d
\Psi_{a+b}
\right].
\end{align}
Likewise, using
\begin{equation}
[B,G]=-\delta_{b,d}\delta(x-y),
\end{equation}
and relabelling $c\to d$, one finds
\begin{equation}
\bar T_3=\bar T_2.
\end{equation}
Thus
\begin{align}
\bar T_2+\bar T_3
=
-2i\sum_{a,b,d\geq1}
w_{a,b}w_{b,d}
\int dx\,
\mathcal R^{(+)}_{b,d}
\left[
\Psi^\dagger_a
\Psi^\dagger_{b+d}
\Psi_d
\Psi_{a+b}
\right].
\end{align}
We now perform the simultaneous relabelling
\begin{equation}
a\to d,
\qquad
b\to a,
\qquad
d\to b.
\end{equation}
It gives
\begin{align}
w_{a,b}
w_{b,d}
&\longrightarrow
w_{a,d}w_{a,b},
\\
\mathcal R^{(+)}_{b,d}
&\longrightarrow
\mathcal R^{(+)}_{a,b},
\end{align}
Moreover,
\begin{align}
\Psi_d^\dagger
\Psi_{a+b}^\dagger
\Psi_b
\Psi_{a+d}
=
\Psi_{a+b}^\dagger
\Psi_d^\dagger
\Psi_{a+d}
\Psi_b,
\end{align}
hence

\begin{align}
\bar T_2+\bar T_3
=
-2i\sum_{a,b,d\geq1}
w_{a,b}w_{a,d}
\int dx\,
\mathcal R^{(+)}_{a,b}
\left[
\Psi^\dagger_{a+b}
\Psi^\dagger_d
\Psi_{a+d}
\Psi_b
\right].
\end{align}

Introducing 
\begin{equation}
Q_{a+d,b}^{a+b,d}(x)
=
\Psi^\dagger_{a+b}(x)
\Psi^\dagger_d(x)
\Psi_{a+d}(x)
\Psi_b(x),
\end{equation}
we finally obtain
\begin{align}
B_2
=
2i\sum_{a,b,d\geq1}
w_{a,b}w_{a,d}
\int dx\,
\left(
\mathcal R^{(-)}_{a,d}
-
\mathcal R^{(+)}_{a,b}
\right)
Q_{a+d,b}^{a+b,d}(x).
\label{eq:second-mixed-block}
\end{align}

Taking into account the derivatives acting on the contracted fields,
the effective differential operators are
\begin{align}
\mathcal R^{(-)}_{a,d}
&\longrightarrow
-\frac{1}{a+d}\partial_{a+d}
+\frac{1}{a}
\left(\partial_{a+b}+\partial_b\right)
+\frac{1}{d}\partial_d,
\\
\mathcal R^{(+)}_{a,b}
&\longrightarrow
\frac{1}{a+b}\partial_{a+b}
-\frac{1}{a}
\left(\partial_d+\partial_{a+d}\right)
-\frac{1}{b}\partial_b.
\end{align}
Combining the two contributions, we obtain
\begin{align}
\mathcal R^{(-)}_{a,d}
-
\mathcal R^{(+)}_{a,b}
\rightsquigarrow
{}&
-\frac{1}{a+d}\partial_{a+d}
+\frac{1}{d}\partial_d
-\frac{1}{a+b}\partial_{a+b}
+\frac{1}{b}\partial_b
\nonumber\\
&+
\frac{1}{a}
\left(
\partial_{a+b}
+\partial_b
+\partial_{a+d}
+\partial_d
\right).
\label{eq:B2-effective-R}
\end{align}

Since the terms proportional to $1/a$ form a total derivative, we can omit it. Hence,
modulo total derivatives,
\begin{align}
B_2
\simeq
2i\sum_{a,b,d\geq1}
w_{a,b}w_{a,d}
\int dx\,
\left[
-\frac{1}{a+d}\partial_{a+d}
+\frac{1}{d}\partial_d
-\frac{1}{a+b}\partial_{a+b}
+\frac{1}{b}\partial_b
\right]
Q_{a+d,b}^{a+b,d}(x).
\label{eq:B2-single-reduced}
\end{align}

We finally consider the third mixed block
\begin{equation}
B_3=T_5+\bar T_1.
\end{equation}

We first consider
\begin{equation}
T_5=DF[A,G]BC.
\end{equation}

The contribution $T_5$ receives one term from each of the three
pieces in $\mathcal R^{(-)}_{c,d}J^\dagger_{c,d}$. Using
\begin{equation}
[A,G]
=
[\Psi^\dagger_{a+b}(x),\Psi_{c+d}(y)]
=
-\delta_{a+b,c+d}\delta(x-y),
\end{equation}
we also have
\begin{equation}
[A,\partial_yG]
=
-\delta_{a+b,c+d}\partial_y\delta(x-y).
\end{equation}
Therefore
\begin{align}
T_5
={}&
i\sum_{a,b,c,d\geq1}
w_{a,b}w_{c,d}
\delta_{a+b,c+d}
\int dx\,dy\,\Bigg[
\frac{1}{c+d}
\Psi_c^\dagger(y)\Psi_d^\dagger(y)
\partial_y\delta(x-y)
\Psi_a(x)\Psi_b(x)
\nonumber\\
&\hspace{28mm}
-\frac{1}{c}
(\partial_y\Psi_c^\dagger(y))
\Psi_d^\dagger(y)
\delta(x-y)
\Psi_a(x)\Psi_b(x)
\nonumber\\
&\hspace{28mm}
-\frac{1}{d}
\Psi_c^\dagger(y)
(\partial_y\Psi_d^\dagger(y))
\delta(x-y)
\Psi_a(x)\Psi_b(x)
\Bigg].
\end{align}
Using
\begin{equation}
\int dy\,f(y)\partial_y\delta(x-y)
=
-\partial_x f(x),
\end{equation}
and setting 
\begin{equation}
t=a+b=c+d,
\end{equation}
we obtain
\begin{align}
T_5
=
-i\sum_{\substack{a,b,c,d\geq1\\a+b=c+d}}
w_{a,b}w_{c,d}
\int dx\,\Bigg[
&
\frac{1}{t}
\partial_x
\left(
\Psi_c^\dagger\Psi_d^\dagger
\right)
\Psi_a\Psi_b
+
\frac{1}{c}
(\partial_x\Psi_c^\dagger)
\Psi_d^\dagger\Psi_a\Psi_b
+
\frac{1}{d}
\Psi_c^\dagger
(\partial_x\Psi_d^\dagger)
\Psi_a\Psi_b
\Bigg].
\end{align}

We next consider
\begin{equation}
\bar T_1=AB[C,D]FG,
\end{equation}
with
\begin{equation}
[C,D]
=
[\Psi_{a+b}(x),\Psi^\dagger_{c+d}(y)]
=
\delta_{a+b,c+d}\delta(x-y),
\end{equation}
we find
\begin{align}
\bar T_1
=
-i\sum_{\substack{a,b,c,d\geq1\\a+b=c+d}}
w_{a,b}w_{c,d}
\int dx\,\Bigg[
&
\frac{1}{t}
\Psi_a^\dagger\Psi_b^\dagger
\partial_x(\Psi_c\Psi_d)
+
\frac{1}{c}
\Psi_a^\dagger\Psi_b^\dagger
(\partial_x\Psi_c)\Psi_d
+
\frac{1}{d}
\Psi_a^\dagger\Psi_b^\dagger
\Psi_c(\partial_x\Psi_d)
\Bigg].
\end{align}
Performing the simultaneous relabelling
\begin{equation}
(a,b)\leftrightarrow(c,d),
\end{equation}
we obtain
\begin{align}
\bar T_1
=
-i\sum_{\substack{a,b,c,d\geq1\\a+b=c+d}}
w_{a,b}w_{c,d}
\int dx\,
\left[
\frac{1}{t}
\left(
\partial_a+\partial_b
\right)
+
\frac{1}{a}\partial_a
+
\frac{1}{b}\partial_b
\right]
 Q_{a,b}^{c,d},
\label{eq:Tbar1-single}
\end{align}
where $Q_{a,b}^{c,d}=\Psi^\dagger_c\Psi^\dagger_d\Psi_a\Psi_b.$

Therefore, the third block is 
\begin{align}
B_3
=
-i\sum_{\substack{a,b,c,d\geq1\\a+b=c+d}}
w_{a,b}w_{c,d}
\int dx\,\Bigg[
&
\frac{1}{t}
\left(
\partial_a+\partial_b+\partial_c+\partial_d
\right)+\frac{1}{a}\partial_a
+\frac{1}{b}\partial_b
+\frac{1}{c}\partial_c
+\frac{1}{d}\partial_d
\Bigg]
Q_{a,b}^{c,d}.
\label{eq:B3-single}
\end{align}
Since the term proportional to $1/t$ is a total derivative, we can omit it. Hence, modulo total derivatives
\begin{align}
B_3
\simeq
-i\sum_{\substack{a,b,c,d\geq1\\a+b=c+d}}
w_{a,b}w_{c,d}
\int dx\,
\left[
\frac{1}{a}\partial_a
+\frac{1}{b}\partial_b
+\frac{1}{c}\partial_c
+\frac{1}{d}\partial_d
\right]
Q^{c,d}_{a,b}.
\label{eq:B3-single-reduced}
\end{align}

\paragraph{Sum of $B_1+B_2+B_3$.}

We now combine the three quartic contributions
\begin{equation}
B_1+B_2+B_3
\end{equation}
where $B_1$ is \eqref{eq:B1-expanded}, $B_2$ is \eqref{eq:B2-single-reduced} and $B_3$ is \eqref{eq:B3-single-reduced}. We notice that in $B_2$ and $B_3$ the fields are already normal ordered, whereas in $B_1$ they are not.

It is then convenient to separate in $B_1$ the contribution that is normal ordered (we refer to it as n.o.) with the one with the dirac-delta (we refer to it as contact term c.t.). Specifically, using
\begin{align}
\Psi_a(x)\Psi^\dagger_d(y)=\Psi^\dagger_d(y)\Psi_a(x)+\delta_{a,d}\delta(x-y)
\end{align}
and
\begin{align}
    \widetilde{Q}^{a+b,d}_{a,b+d}(x,y)={Q}^{a+b,d}_{a,b+d}(x,y)+\delta_{a,d}\delta(x-y)\Psi^\dagger_{a+b}(x)\Psi_{b+d}(y),
\end{align}
we can separate
\begin{align}
    B_1=B_1^{n.o}+B_1^{c.t.}.
    \label{separationofB1}
\end{align}

We want to show that
\begin{equation}
B_1^{n.o.}+B_2+B_3
\end{equation}
is a total derivative.

In this way, the three blocks can now be expressed in terms of the same normal ordered quartic vertex,

\begin{equation}
Q^{p,q}_{r,s}(x)
=
\Psi_p^\dagger(x)\Psi_q^\dagger(x)
\Psi_r(x)\Psi_s(x),
\qquad
p+q=r+s\equiv \tau.
\label{eq:common-quartic}
\end{equation}

Since creation operators commute among themselves, and likewise for
annihilation operators,
\begin{equation}
 Q^{p,q}_{r,s}
=
 Q^{q,p}_{r,s}
=
 Q^{p,q}_{s,r}.
\label{eq:Q-symmetries}
\end{equation}

We further define
\begin{equation}
D=
\partial_p+\partial_q+\partial_r+\partial_s,
\qquad
S=
\frac1p\partial_p+
\frac1q\partial_q+
\frac1r\partial_r+
\frac1s\partial_s.
\label{eq:D-def}
\end{equation}
By construction,
\begin{equation}
D\, Q^{p,q}_{r,s}
=
\partial_x Q^{p,q}_{r,s}.
\end{equation}
\paragraph{First block.}

Modulo total derivatives, the normally ordered part of the first
block is
\begin{align}
B_1^{\rm n.o.}
\simeq
2i\sum_{a,b,d\geq1}
w_{a,b}w_{b,d}
\int dx\,
\Bigg[
&
-\frac{1}{b+d}\partial_{b+d}
+\frac{1}{d}\partial_d
-\frac{1}{a+b}\partial_{a+b}
+\frac{1}{a}\partial_a
\Bigg]
Q_{a,b+d}^{a+b,d}.
\label{eq:B1-normal-reduced}
\end{align}
For $B_1^{\rm n.o.}$ we introduce
\begin{equation}
p=a+b,
\qquad
q=d,
\qquad
r=a,
\qquad
s=b+d.
\label{eq:B1-map}
\end{equation}
They satisfy
\begin{equation}
p+q=r+s=\tau.
\end{equation}
Moreover,
\begin{equation}
b=p-r=s-q.
\end{equation}
Let
\begin{equation}
\xi\equiv p-r=s-q.
\label{eq:xi-def}
\end{equation}
Since $b\geq1$, the first block corresponds to the region
\begin{equation}
\xi>0.
\end{equation}

The product of weights becomes
\begin{align}
w_{a,b}w_{b,d}
=
\sqrt{ab(a+b)}\sqrt{bd(b+d)}
=
b\sqrt{ad(a+b)(b+d)}
=
\xi\sqrt{pqrs}.
\label{eq:B1-weight}
\end{align}

The differential operator in \eqref{eq:B1-normal-reduced} becomes
\begin{equation}
A
\equiv
-\frac{1}{p}\partial_p
+\frac{1}{q}\partial_q
+\frac{1}{r}\partial_r
-\frac{1}{s}\partial_s.
\label{eq:A-def}
\end{equation}
Therefore
\begin{equation}
B_1^{n.o.}
\simeq
2i
\sum_{\substack{p,q,r,s\geq1\\p+q=r+s\\\xi>0}}
\sqrt{pqrs}
\int dx\,
\xi A\,
 Q^{p,q}_{r,s}.
\label{eq:B1-common}
\end{equation}

If we consider \eqref{eq:B1-map}, it may seem that the sum for $p$ and $s$ has to start from $2$. This is already taken into account, in fact since $\xi>0$, also $p>r$ and $s>q$ and, since $r,q>0$, it implies $p,s \ge 2$.

\paragraph{Second block.}

Similarly,
\begin{align}
B_2
\simeq
2i\sum_{a,b,d\geq1}
w_{a,b}w_{a,d}
\int dx\,
\Bigg[
-\frac{1}{a+d}\partial_{a+d}
+\frac{1}{d}\partial_d
-\frac{1}{a+b}\partial_{a+b}
+\frac{1}{b}\partial_b
\Bigg]
 Q^{a+b,d}_{a+d,b}.
\label{eq:B2-before-common}
\end{align}

This time we introduce
\begin{equation}
p=a+b,
\qquad
q=d,
\qquad
r=a+d,
\qquad
s=b.
\label{eq:B2-map}
\end{equation}
Again,
\begin{equation}
p+q=r+s=\tau.
\end{equation}
Moreover,
\begin{equation}
a=p-s=r-q.
\end{equation}
We define
\begin{equation}
\eta\equiv p-s=r-q.
\label{eq:eta-def}
\end{equation}
Since $a\geq1$, this block corresponds to
\begin{equation}
\eta>0.
\end{equation}

The product of weights becomes
\begin{align}
w_{a,b}w_{a,d}
=
\sqrt{ab(a+b)}\sqrt{ad(a+d)}
=
a\sqrt{bd(a+b)(a+d)}
=
\eta\sqrt{pqrs}.
\label{eq:B2-weight}
\end{align}

The corresponding differential operator is
\begin{equation}
B
\equiv
-\frac{1}{p}\partial_p
+\frac{1}{q}\partial_q
-\frac{1}{r}\partial_r
+\frac{1}{s}\partial_s.
\label{eq:B-def}
\end{equation}
Hence
\begin{equation}
B_2
\simeq
2i
\sum_{\substack{p,q,r,s\geq1\\p+q=r+s\\\eta>0}}
\sqrt{pqrs}
\int dx\,
\eta B\,
 Q^{p,q}_{r,s}.
\label{eq:B2-common}
\end{equation}

Similar to before, the range of $p,r\ge 2$ is encountered in $\eta>0$.

\paragraph{Combining the first two blocks.}

Introduce
\begin{equation}
[z]_+\equiv\max(z,0).
\end{equation}
Equations \eqref{eq:B1-common} and \eqref{eq:B2-common} may then be
written over the same complete domain as
\begin{align}
B_1^{n.o.}+B_2
\simeq
2i
\sum_{\substack{p,q,r,s\geq1\\p+q=r+s}}
\sqrt{pqrs}
\int dx\,
\left(
[\xi]_+ A+[\eta]_+B
\right)
 Q^{p,q}_{r,s}.
\label{eq:B12-positive-part}
\end{align}

Although the original changes of variables imply
$p,s\geq2$ in $B_1^{n.o.}$ and $p,r\geq2$ in $B_2$,
these restrictions are already encoded in the positivity conditions.
Indeed,
\begin{equation}
\xi>0
\quad\Longrightarrow\quad
p=r+\xi\geq2,
\qquad
s=q+\xi\geq2,
\end{equation}
whereas
\begin{equation}
\eta>0
\quad\Longrightarrow\quad
p=s+\eta\geq2,
\qquad
r=q+\eta\geq2.
\end{equation}
Therefore, both blocks can be extended to the common domain
$p,q,r,s\geq1$, $p+q=r+s$, by introducing the positive parts
$[\xi]_+$ and $[\eta]_+$.

We now exploit the symmetry under the exchange of the two creation
indices,
\begin{equation}
p\leftrightarrow q.
\end{equation}
Since the summation indices are dummy and
$ Q^{p,q}_{r,s}= Q^{q,p}_{r,s}$, this relabelling
does not change the value of the sum.

Under $p\leftrightarrow q$, using $p+q=r+s$, we have
\begin{align}
\xi=p-r
&\longrightarrow
q-r
=
s-p
=
-\eta,
\\
\eta=p-s
&\longrightarrow
q-s
=
r-p
=
-\xi.
\label{eq:xi-eta-transform}
\end{align}
Moreover,
\begin{align}
A
&\longrightarrow
-\frac{1}{q}\partial_q
+\frac{1}{p}\partial_p
+\frac{1}{r}\partial_r
-\frac{1}{s}\partial_s
=
-B,
\\
B
&\longrightarrow
-\frac{1}{q}\partial_q
+\frac{1}{p}\partial_p
-\frac{1}{r}\partial_r
+\frac{1}{s}\partial_s
=
-A.
\label{eq:AB-transform}
\end{align}

Thus the relabelled coefficient is
\begin{equation}
-[ -\eta ]_+ B
-
[ -\xi ]_+ A.
\end{equation}
Since the original and relabelled expressions give the same sum, we
may replace the coefficient in \eqref{eq:B12-positive-part} by their
average. We obtain
\begin{align}
&
\frac{1}{2}
\Big[
[\xi]_+A
+
[\eta]_+B
-
[-\xi]_+A
-
[-\eta]_+B
\Big]
=
\frac{1}{2}
\Big[
\left([\xi]_+-[-\xi]_+\right)A
+
\left([\eta]_+-[-\eta]_+\right)B
\Big].
\end{align}
Using the elementary identity
\begin{equation}
[z]_+-[-z]_+=z,
\end{equation}
this reduces to
\begin{equation}
\frac{1}{2}
\left(
\xi A+\eta B
\right).
\end{equation}
The factor $1/2$ cancels the factor $2$ in
\eqref{eq:B12-positive-part}, and therefore
\begin{equation}
B_1^{n.o.}+B_2
\simeq
i
\sum_{\substack{p,q,r,s\geq1\\p+q=r+s}}
\sqrt{pqrs}
\int dx\,
\left(
\xi A+\eta B
\right)
 Q^{p,q}_{r,s}.
\label{eq:B12-symmetrised}
\end{equation}

\paragraph{Simplification of the differential operator.}

Using
\begin{equation}
\xi=p-r,
\qquad
\eta=p-s,
\end{equation}
together with \eqref{eq:A-def} and \eqref{eq:B-def}, we find
\begin{align}
\xi A+\eta B
={}&
-\frac{\xi+\eta}{p}\partial_p
+\frac{\xi+\eta}{q}\partial_q
+\frac{\xi-\eta}{r}\partial_r
+\frac{\eta-\xi}{s}\partial_s.
\label{eq:xiAetaB-first}
\end{align}

We now use
\begin{align}
\xi+\eta
&=
(p-r)+(p-s)
=
2p-(r+s)
=
2p-(p+q)
=
p-q,
\\
\xi-\eta
&=
(p-r)-(p-s)
=
s-r,
\\
\eta-\xi
&=
r-s.
\end{align}
Hence
\begin{align}
\xi A+\eta B
={}&
-\frac{p-q}{p}\partial_p
+\frac{p-q}{q}\partial_q
+\frac{s-r}{r}\partial_r
+\frac{r-s}{s}\partial_s.
\label{eq:xiAetaB-second}
\end{align}

Since
\begin{equation}
\tau=p+q=r+s,
\end{equation}
the four coefficients may be rewritten as
\begin{align}
-\frac{p-q}{p}
&=
\frac{\tau}{p}-2,
&
\frac{p-q}{q}
&=
\frac{\tau}{q}-2,
\\
\frac{s-r}{r}
&=
\frac{\tau}{r}-2,
&
\frac{r-s}{s}
&=
\frac{\tau}{s}-2.
\end{align}
Therefore
\begin{align}
\xi A+\eta B
={}&
\tau
\left(
\frac{1}{p}\partial_p
+\frac{1}{q}\partial_q
+\frac{1}{r}\partial_r
+\frac{1}{s}\partial_s
\right)
-
2
\left(
\partial_p+\partial_q+\partial_r+\partial_s
\right).
\end{align}
In terms of \eqref{eq:D-def},
\begin{equation}
\xi A+\eta B=\tau S-2D.
\label{eq:xiAetaB-final}
\end{equation}

Consequently,
\begin{equation}
B_1^{n.o.}+B_2
\simeq
i
\sum_{\substack{p,q,r,s\geq1\\p+q=r+s}}
\sqrt{pqrs}
\int dx\,
\left(
\tau S-2D
\right)
 Q^{p,q}_{r,s}.
\label{eq:B12-final}
\end{equation}

\paragraph{Third block.}

The third block is
\begin{align}
B_3
\simeq
-i
\sum_{\substack{a,b,c,d\geq1\\a+b=c+d}}
w_{a,b}w_{c,d}
\int dx\,
\left[
\frac{1}{a}\partial_a
+\frac{1}{b}\partial_b
+\frac{1}{c}\partial_c
+\frac{1}{d}\partial_d
\right]
 Q_{a,b}^{c,d}.
\label{eq:B3-start-common}
\end{align}

We introduce
\begin{equation}
p=c,
\qquad
q=d,
\qquad
r=a,
\qquad
s=b.
\label{eq:B3-map}
\end{equation}
Then
\begin{equation}
p+q=r+s=\tau.
\end{equation}
Moreover,
\begin{align}
w_{a,b}w_{c,d}
&=
\sqrt{ab(a+b)}
\sqrt{cd(c+d)}
\nonumber\\
&=
\sqrt{rs\tau}\sqrt{pq\tau}
\nonumber\\
&=
\tau\sqrt{pqrs}.
\label{eq:B3-weight-common}
\end{align}
The differential operator becomes precisely $S$,
\begin{equation}
\frac{1}{a}\partial_a
+\frac{1}{b}\partial_b
+\frac{1}{c}\partial_c
+\frac{1}{d}\partial_d
=
S.
\end{equation}
Therefore
\begin{equation}
B_3
\simeq
-i
\sum_{\substack{p,q,r,s\geq1\\p+q=r+s}}
\sqrt{pqrs}
\int dx\,
\tau S\, Q^{p,q}_{r,s}.
\label{eq:B3-common-final}
\end{equation}

\paragraph{Final cancellation.}

Combining \eqref{eq:B12-final} and
\eqref{eq:B3-common-final}, we obtain
\begin{align}
B_1^{n.o.}+B_2+B_3
\simeq
i
\sum_{\substack{p,q,r,s\geq1\\p+q=r+s}}
\sqrt{pqrs}
\int dx\,
\Big[
(\tau S-2D)-\tau S
\Big]
 Q^{p,q}_{r,s}.
\end{align}
The terms proportional to $\tau  S$ cancel identically, leaving
\begin{align}
B_1^{n.o.}+B_2+B_3
\simeq
-2i
\sum_{\substack{p,q,r,s\geq1\\p+q=r+s}}
\sqrt{pqrs}
\int dx\,
D\, Q^{p,q}_{r,s}.
\end{align}
Using
\begin{equation}
D\, Q^{p,q}_{r,s}
=
\partial_x Q^{p,q}_{r,s},
\end{equation}
we finally find
\begin{equation}
{
B_1^{n.o.}+B_2+B_3
\simeq
-2i
\sum_{\substack{p,q,r,s\geq1\\p+q=r+s}}
\sqrt{pqrs}
\int dx\,
\partial_x Q^{p,q}_{r,s}.
}
\label{eq:mixed-single-total-derivative}
\end{equation}
Therefore, assuming the usual boundary conditions,
\begin{equation}
{
B_1^{n.o.}+B_2+B_3=0.
}
\end{equation}

\paragraph{Contact term from the normal ordering of $B_1$.}

We consider the separation of $B_1$ into the normal ordered part and the contact term given in \eqref{separationofB1}. The normally ordered part of this expression has already been included
in the cancellation $B_1^{\rm n.o.}+B_2+B_3=0.$

We now study the remaining contact contribution $B_1^{\rm ct}$, 

\begin{align}
B_1^{\rm c.t.}
\simeq
2i\sum_{a,b,d\geq1}
w_{a,b}w_{b,d}
\int dx\,
\Bigg[
&
\underbrace{-\frac{1}{b+d}\partial_{b+d}}_{(1)}
+\underbrace{\frac{1}{d}\partial_d}_{(2)}
\underbrace{-\frac{1}{a+b}\partial_{a+b}}_{(3)}
+\underbrace{\frac{1}{a}\partial_a}_{(4)}
\Bigg]
\delta_{a,d}\delta(x-y)\Psi^\dagger_{a+b}(x)\Psi_{b+d}(y).
\label{eq:B1-ct}
\end{align}
For simplicity, we have numbered as $(1)$ to $(4)$ each terms of this expression.

Since fields  are operator-valued distributions,
as usual products of contractions at the same point require an ultraviolet
prescription. 
We therefore introduce a translationally invariant and symmetric
ultraviolet regularisation of the canonical commutator,
\begin{equation}
[\Psi_m^{(\Lambda)}(x),
\Psi_n^{(\Lambda)\dagger}(y)]
=
\delta_{m,n}\,
K_\Lambda(x-y),
\label{eq:regulated-CCR}
\end{equation}
where
\begin{equation}
K_\Lambda(z)=K_\Lambda(-z),
\qquad
K_\Lambda(z)\xrightarrow{\Lambda\to\infty}\delta(z).
\label{eq:regulator-properties}
\end{equation}
Such a regularisation can be obtained, for instance, by introducing
a symmetric momentum-space cutoff.

At finite $\Lambda$, $K_\Lambda(x-y)$ is an ordinary real function that approximates the dirac delta, so
products such as
$K_\Lambda(x-y)^2$ and
$K_\Lambda(x-y)\partial_xK_\Lambda(x-y)$
are well defined.
Importantly, the argument below does not depend on the detailed form
of $K_\Lambda$. It uses only translation invariance and the symmetry
$K_\Lambda(z)=K_\Lambda(-z)$. Therefore, the cancellation is not tied
to a particular choice of regulator, but holds for the whole class
of symmetric translationally invariant ultraviolet regularisations.

The terms (1) and (2) of \eqref{eq:B1-ct}
originate from $T_1+T_2$ and have the bilocal ordering
\begin{equation}
\Psi^\dagger_{a+b}(x)\Psi_a(x)
\Psi^\dagger_d(y)\Psi_{b+d}(y),
\end{equation}
whereas (3) and (4) originate from
$\bar T_4+\bar T_5$ and have the opposite assignment of positions,
\begin{equation}
\Psi^\dagger_{a+b}(y)\Psi_a(y)
\Psi^\dagger_d(x)\Psi_{b+d}(x).
\end{equation}

Let us first consider the term (1) .
The second contraction is
\begin{equation}
[\Psi_a(x),\Psi_d^\dagger(y)]
=
\delta_{a,d}K_\Lambda(x-y).
\end{equation}
Together with the contraction already present in $T_1+T_2$, this gives
\begin{align}
B_{1,\Lambda}^{(1)}
=
-2i\sum_{a,b\geq1}
w_{a,b}^2
\int dx\,dy\,
\frac{1}{\rho}
K_\Lambda(x-y)^2
\Psi_\rho^\dagger(x)
\partial_y\Psi_\rho(y),
\label{eq:B1-contact-1}
\end{align}
where
\begin{equation}
\rho\equiv a+b.
\end{equation}
Indeed, the second contraction sets $d=a$, so that
$b+d=a+b=\rho$ and
\begin{equation}
w_{a,b}w_{b,d}\big|_{d=a}
=
w_{a,b}^2.
\end{equation}

The term (2) contains instead a derivative on the
field involved in the second contraction. Using
\begin{align}
[\Psi_a(x),\partial_y\Psi_d^\dagger(y)]
&=
\delta_{a,d}\partial_yK_\Lambda(x-y)
=
-\delta_{a,d}\partial_xK_\Lambda(x-y),
\end{align}
we obtain
\begin{align}
B_{1,\Lambda}^{(2)}
=
-2i\sum_{a,b\geq1}
w_{a,b}^2
\int dx\,dy\,
\frac{1}{a}
K_\Lambda(x-y)
\partial_xK_\Lambda(x-y)
\Psi_\rho^\dagger(x)\Psi_\rho(y).
\label{eq:B1-contact-2}
\end{align}

We now consider the terms (3) and (4). The term (3) gives
\begin{align}
B_{1,\Lambda}^{(3)}
=
-2i\sum_{a,b\geq1}
w_{a,b}^2
\int dx\,dy\,
\frac{1}{\rho}
K_\Lambda(x-y)^2
\left(\partial_y\Psi_\rho^\dagger(y)\right)
\Psi_\rho(x).
\label{eq:B1-contact-3}
\end{align}

Finally, using
\begin{align}
[\partial_y\Psi_a(y),\Psi_d^\dagger(x)]
=
\delta_{a,d}\partial_yK_\Lambda(y-x)
=
-\delta_{a,d}\partial_xK_\Lambda(x-y),
\end{align}
the term (4) gives
\begin{align}
B_{1,\Lambda}^{(4)}
=
-2i\sum_{a,b\geq1}
w_{a,b}^2
\int dx\,dy\,
\frac{1}{a}
K_\Lambda(x-y)
\partial_xK_\Lambda(x-y)
\Psi_\rho^\dagger(y)\Psi_\rho(x).
\label{eq:B1-contact-4}
\end{align}

It is convenient to introduce the symmetric bilocal operator
\begin{equation}
H_\rho(x,y)
\equiv
\Psi_\rho^\dagger(x)\Psi_\rho(y)
+
\Psi_\rho^\dagger(y)\Psi_\rho(x),
\label{eq:H-bilocal}
\end{equation}
which satisfies
\begin{equation}
H_\rho(x,y)=H_\rho(y,x).
\end{equation}
Adding \eqref{eq:B1-contact-1}, \eqref{eq:B1-contact-2}, \eqref{eq:B1-contact-3}, \eqref{eq:B1-contact-4},
the full contact contribution can be written as
\begin{align}
B_{1,\Lambda}^{\rm ct}
=
-2i\sum_{a,b\geq1}
w_{a,b}^2
\int dx\,dy\,\Bigg[
&
\frac{1}{\rho}
K_\Lambda(x-y)^2
\partial_yH_\rho(x,y)
\nonumber\\
&+
\frac{1}{a}
K_\Lambda(x-y)
\partial_xK_\Lambda(x-y)
H_\rho(x,y)
\Bigg].
\label{eq:B1-contact-cutoff}
\end{align}

We now show that both terms in
\eqref{eq:B1-contact-cutoff} vanish already at finite cutoff.

For the second term, under the exchange $x\leftrightarrow y$,
$K_\Lambda(x-y)$ is even, whereas
$\partial_xK_\Lambda(x-y)$ is odd.
Since $H_\rho(x,y)$ is symmetric, the full integrand is antisymmetric.
Therefore
\begin{equation}
\int dx\,dy\,
K_\Lambda(x-y)
\partial_xK_\Lambda(x-y)
H_\rho(x,y)
=
0.
\label{eq:B1-contact-odd}
\end{equation}

For the first term, using the symmetry of both
$K_\Lambda(x-y)^2$ and $H_\rho(x,y)$ under
$x\leftrightarrow y$, we have
\begin{align}
\int dx\,dy\,
 K_\Lambda(x-y)^2
\partial_yH_\rho(x,y)
=
\frac{1}{2}
\int dx\,dy\,
 K_\Lambda(x-y)^2
(\partial_x+\partial_y)H_\rho(x,y).
\label{eq:B1-contact-average}
\end{align}
After integration by parts,
\begin{align}
&
\int dx\,dy\,
 K_\Lambda(x-y)^2
(\partial_x+\partial_y)H_\rho(x,y)
\nonumber
=
-\int dx\,dy\,
(\partial_x+\partial_y)
 K_\Lambda(x-y)^2
H_\rho(x,y).
\end{align}
Since the regulator is translationally invariant,
$ K_\Lambda$ depends only on $x-y$, and therefore
\begin{equation}
(\partial_x+\partial_y)
K_\Lambda(x-y)^2
=
0.
\end{equation}
Hence
\begin{equation}
\int dx\,dy\,
K_\Lambda(x-y)^2
\partial_yH_\rho(x,y)
=
0.
\label{eq:B1-contact-even}
\end{equation}

We thus conclude that the contact contribution vanishes identically
for every finite value of the cutoff,
\begin{equation}
{
B_{1,\Lambda}^{\rm ct}=0.
}
\end{equation}
The removal of the regulator is therefore immediate,
\begin{equation}
{
B_1^{\rm ct}
=
\lim_{\Lambda\to\infty}
B_{1,\Lambda}^{\rm ct}
=
0.
}
\label{eq:B1-contact-zero}
\end{equation}

Together with
\begin{equation}
B_1^{\rm n.o.}+B_2+B_3=0,
\end{equation}
this gives
\begin{equation}
{
B_1+B_2+B_3=0.
}
\end{equation}
Therefore, the mixed sector vanishes,
\begin{equation}
[C^{(+)},G^{(-)}]
+
[C^{(-)},G^{(+)}]
=
0.
\end{equation}

\section{Bethe Ansatz}
\label{betheansatzsection}
Let us focus on the mass-$2$ sector, which consists of a superposition of a wave function with two (bosonic) mass-$1$ particles and one (bosonic) mass-$2$ particle. The energy eigenstate in this sector is given by \cite{Kozlowski:2016too}
\begin{eqnarray}
|f\rangle = \int_{x_1 < x_2} dx_1 \, dx_2 \, f_2 \begin{pmatrix}1&1\\x_1&x_2\end{pmatrix} \, \psi_1^{\dagger}(x_1)\, \psi_1^{\dagger}(x_2)|0\rangle + \int_{-\infty}^{\infty} dx \, f_1 \begin{pmatrix}2\\x\end{pmatrix} \, \psi_2^{\dagger}(x)\, |0\rangle.
\end{eqnarray} 
We now place this wave function in a finite box $\Big[-\frac{L}{2},\frac{L}{2}\Big]$, and imposing periodic boundary condition. Namely, we require that both
\begin{eqnarray}
\langle 0|\psi_1(y_1) \, \psi_1(y_2)|f\rangle 
\end{eqnarray}
and 
\begin{eqnarray}
\langle 0|\psi_2(y_1)|f\rangle 
\end{eqnarray}
attain the same value at $y_1 = - \frac{L}{2}$ and at $y_1 = \frac{L}{2}$. 

A small calculation using the commutation relations from \cite{Kozlowski:2016too} leads to the two conditions
\begin{eqnarray}
f_2 \begin{pmatrix}1&1\\-\frac{L}{2}&x_2\end{pmatrix}=f_2 \begin{pmatrix}1&1\\\frac{L}{2}&x_2\end{pmatrix}, \qquad f_1 \begin{pmatrix}2\\-\frac{L}{2}\end{pmatrix}=f_1 \begin{pmatrix}2\\\frac{L}{2}\end{pmatrix}\label{in}.
\end{eqnarray}
Using now the explicit expression of the symmetric wave functions from \cite{Kozlowski:2016too} , this amounts to
\begin{eqnarray}
e^{i u_1 L} = S_{21}, \qquad e^{-i u_2 L} = S_{21},\label{Bethe}
\end{eqnarray}
which automaticaly solves also the second condition in (\ref{in}). We can further reduce this to one single equation, if we notice that we need to require
\begin{eqnarray}
u_2 = \frac{2\pi n }{L} - u_1, \qquad n \in \mathbbmss{Z},
\end{eqnarray}
leading to the Bethe equation
\begin{eqnarray}
e^{i u_1 L} = -\frac{P\Big(\frac{2i\pi n }{L} - 2i u_1\Big)}{P\Big(-\frac{2i\pi n }{L} + 2i u_1\Big)},
\label{BE}
\end{eqnarray}
with $P(v) = v^3 + 12 \gamma v -4\beta^2$ as in \cite{Kozlowski:2016too}. We see that it is consistent to try to solve this equation with real $u_1$, which corresponds to the unbound spectrum. 

 Complex solutions are also
possible and are associated with bound states. Requiring the energy to be real, they occur in complex-conjugate pairs. In the following, we restrict our analysis to the unbound spectrum.

In this section, we first provide a perturbative solution. Later, we present the exact one.

We solve the Bethe equation perturbatively around $\beta^2=0$ (the free point $S_{21}=1)$. We write the solution in a series expansion
\begin{eqnarray}
u_1 = u_1^{(0)} + \beta^2 u_1^{(1)} + \beta^4 u_1^{(2)}+\beta^6  u_1^{(3)}... 
\end{eqnarray}  
and the fulfillment of the equation requires
\begin{eqnarray}
&&u_1^{(0)} = \frac{2\pi m }{L}, \qquad m \in \mathbbmss{Z},\nonumber\\
&&u_1^{(1)} = \frac{L^2}{(2m - n)^3 \pi^3 + 3L^2 \pi \gamma (n - 2m)},\nonumber\\
&&u_1^{(2)} = \frac{-3L^5 \pi^2(n-2m)^2 + 3L^7 \gamma}{\Big[(2m - n)^3 \pi^3 + 3L^2 \pi \gamma (n - 2m)\Big]^3}.
\end{eqnarray}
By a suitable choice of variables, we can push this analysis further. We define
\begin{eqnarray}
\zeta \equiv u_1^{(0)} - \frac{n \pi}{L}.  \label{zeta}  
\end{eqnarray}
One then finds that 
\begin{eqnarray}
u_1^{(M)} = \frac{p_1^{(M)}}{[- L^3 \zeta (\zeta^2 - 3 \gamma)]^{2M-1}},
\end{eqnarray}
where $p_1^{(M)}$ is a polynomial in $\gamma$ and $\zeta$ which can be calculated recursively. The first few orders are
\begin{eqnarray}
&&p_1^{(1)} = - L^2,\nonumber\\
&&p_1^{(2)} =  3 L^7 (\zeta^2 - \gamma),\nonumber\\
&&p_1^{(3)} = - \frac{L^{12}}{12}\Big(-216 \gamma^2 + 9 \gamma \zeta^2 (36 + L^2 \gamma) - 6\zeta^4 (30+L^2 \gamma)+L^2 \zeta^6\Big),\nonumber\\
&&p_1^{(4)} = L^{17} \Big(-135 \gamma^3 + 9 \gamma^2 \zeta^2 (31 + L^2 \gamma) - 3 \gamma \zeta^4 (77 + 5 L^2 \gamma) + 7 \zeta^6 (13+L^2\gamma)-L^2 \zeta^8\Big).\end{eqnarray}
Using Mathematica, we computed the first 15 terms of this expansion.

In the special case of $\gamma=0$ (zero rest mass), we can go a bit further:
\begin{eqnarray}
&&u_1^{(0)} = \frac{2\pi m }{L}, \qquad m \in \mathbbmss{Z},\nonumber\\
&&u_1^{(1)} = \frac{L^2}{(2m - n)^3 \pi^3},\nonumber\\
&&u_1^{(2)} = \frac{-3L^5}{(2m - n)^7 \pi^7}\nonumber\\
&&u_1^{(3)} = \frac{-L^8\Big(\pi^2(n-2m)^2-180\Big)}{12(2m - n)^{11} \pi^{11}}.
\end{eqnarray}
Using the variable $\zeta$, we obtain, in this case, the first 20 orders of the expansion,
\begin{eqnarray}
u_1^{(M)} = -\frac{p_1^{(M)}}{(L \zeta )^{6M-3}},
\end{eqnarray}
\begin{eqnarray}
&&p_1^{(1)} = - L^2,\nonumber\\
&&p_1^{(2)} =  3 L^7 \zeta^2,\nonumber\\
&&p_1^{(3)} = \frac{L^{12}\zeta^4 }{12}\Big(-180+L^2 \zeta^2\Big),\nonumber\\
&&p_1^{(4)} = L^{17}\zeta^6 \Big(91 - L^2 \zeta^2\Big),\nonumber\\
&&p_1^{(5)} = -\frac{L^{22}\zeta^8}{80}\Big(48960 - 800 L^2 \zeta^2 + L^4 \zeta^4\Big).\label{pexpansions}\end{eqnarray}

The values are needed to compute the energy
\begin{eqnarray}
E = u_1^2 + \Big(u_1 - \frac{2\pi n}{L}\Big)^2 + 2 \gamma,
\end{eqnarray}
but the wave functions can be simplified just using recursively (\ref{Bethe}), at least the mass-$1$ part. The overall normalisation is also relatively easy to compute and not particularly illuminating. 

\subsection{Exact solution for $\gamma=0$}

We consider the equation \eqref{BE} and rewrite it as 
\begin{equation}
    \beta^2=\frac{2 (L u_1-\pi  n)^3 \tan \left(\frac{L u_1}{2}\right)}{L^3}.
    \label{betheequationgamma0}
\end{equation}
We consider $u_1(\beta)=t(\beta)+\zeta+\frac{n \pi}{L}$, where the non-trivial dependence on $\beta$ is confined in a function $t(\beta)$ and $\zeta$ is constant. We get a condition on $\zeta$ from the expansion at zero order in $\beta$. In particular, from \eqref{betheequationgamma0} and by requiring $t(0)=0$, we get 
(if we decide to write $u_0$ in place of  $u_1^{(0)}$ for brevity)
\begin{equation}
    0={2 (L u_0-\pi  n)^3 \tan \left(\frac{L u_0}{2}\right)} = 2 \zeta^3 L^3 \tan \left(\frac{L u_0}{2}\right).
\end{equation}
We consider $\zeta\neq0$, $\tan \left(\frac{L u_0}{2}\right)=0$ and so $n \pi+L \zeta=2 m \pi$,
\begin{equation}
    \beta^2=2 (\zeta +t)^3 \tan(\frac{L t}{2})
    \label{beta2gamma0}.
\end{equation}
We define 
\begin{eqnarray}
  F(t)=  2 (\zeta +t)^3 \tan(\frac{L t}{2})
\end{eqnarray}
so 
\begin{equation}
    u_1=\frac{n \pi}{L}+\zeta+F^{-1}(\beta^2).
\end{equation}

The inverse function can be expanded around $\beta=0$ by means of the Lagrange--B\"urmann inversion formula with starting point $0$. We write
\begin{equation}
u_1(\beta)
=\frac{n\pi}{L}+\zeta
+\sum_{k=1}^{\infty}c_k\beta^{2k},
\end{equation}
the coefficients can be written as
\begin{equation}
c_k
=
\frac{P_k\left(L^2\zeta^2\right)}
{L^k\zeta^{4k-1}},
\end{equation}
where $P_k(x)$ is a polynomial of degree
\begin{equation}
\deg P_k=\left\lfloor\frac{k-1}{2}\right\rfloor
\end{equation}
and 
\begin{equation}
P_k(L^2)
=
\frac{1}{k}
\left[t^{k-1}\right]
\left[
\frac{L t}
{2(1+t)^3\tan\left(\frac{Lt}{2}\right)}
\right]^k.
\label{eq:ck-Lagrange}
\end{equation}
Here, $\left[t^m\right]f(t)$ denotes the coefficient of $t^m$ in the expansion of $f(t)$.

The first polynomials are
\begin{align}
&P_1(x)=1,\
&&P_2(x)=-3,\
&&P_3(x)=15-\frac{x}{12},\
&&P_4(x)=-91+x,\
&&P_5(x)=612-10x+\frac{x^2}{80},
\end{align}
\begin{align}
&P_6(x)=-4389+95x-\frac{23x^2}{80},
\end{align}
which coincides with \eqref{pexpansions}.

Therefore, the expansion of the solution starts as
\begin{align}
u_1(\beta)
={}&\frac{n\pi}{L}+\zeta
+\frac{\beta^2}{L\zeta^3}
-\frac{3\beta^4}{L^2\zeta^7}
+\frac{180-L^2\zeta^2}{12L^3\zeta^{11}}\beta^6
+\frac{-91+L^2\zeta^2}{L^4\zeta^{15}}\beta^8
+\\
&\frac{48960-800L^2\zeta^2+L^4\zeta^4}
{80L^5\zeta^{19}}\beta^{10}
+\mathcal{O}(\beta^{12}).
\end{align}

It is useful to reorganize the solution as an expansion in the dimensionless variable
\begin{equation}
x=L\zeta.
\end{equation}
We also introduce
\begin{equation}
q=\frac{\beta^2}{L\zeta^4},
\quad s = \frac{tL}{x}, \quad 
u_1(\beta)=\frac{n\pi}{L}+\zeta+\frac{x s}{L}.
\end{equation}
By starting from \eqref{beta2gamma0}, the implicit equation for $u_1$ then becomes
\begin{equation}
q=
\frac{2}{x}(1+s)^3 
\tan\left(\frac{x s}{2}\right).
\label{eq:implicit-s}
\end{equation}

We invert this equation again using the same technique as earlier, and expand $s(q,x)$ as
\begin{equation}
s(q,x)=\sum_{j=1}^{\infty}p_j(x)q^j,
\end{equation}
where $p_j(x)$ is an even polynomial,
\begin{equation}
p_j(x)=
\sum_{k=0}^{\lfloor (j-1)/2\rfloor}
c_{j,k}x^{2k}.
\end{equation}
Therefore,
\begin{equation}
u_1(\beta)
=
\frac{n\pi}{L}+\zeta+
\sum_{j=1}^{\infty}
\frac{p_j(L\zeta)}
{L^j\zeta^{4j-1}}
\beta^{2j}.
\label{u335u1expansion}
\end{equation}

By the Lagrange--B\"urmann inversion formula, the polynomials $p_j(x)$ are given by
\begin{equation}
p_j(x)
=
\frac{1}{j}
\left[z^{j-1}\right]
\left[
\frac{xz/2}
{(1+z)^3\tan(xz/2)}
\right]^j.
\label{eq:pj-Lagrange}
\end{equation}
Here, $[z^m]f(z)$ denotes the coefficient of $z^m$ in the Taylor expansion of $f(z)$.

To obtain an explicit expression for the coefficient of each power of $x$, let
\begin{equation}
A_{j,k}
=
\left[y^{2k}\right](y\cot y)^j.
\end{equation}
Then
\begin{equation}
{
c_{j,k}
=
\frac{(-1)^{j-1}}{j4^k}
\binom{4j-2k-2}{j-2k-1}
A_{j,k}
}
\label{eq:cjk}
\end{equation}
for $0\leq k\leq \lfloor (j-1)/2\rfloor$.

Thus, for each fixed power $x^{2k}$, the coefficient is a binomial coefficient multiplied by a polynomial in the expansion index $j$. 

For $k=0$, since $A_{j,0}=1$, one finds
\begin{equation}
{
[x^0]p_j(x)
=
\frac{(-1)^{j-1}}{j}
\binom{4j-2}{j-1}
=
3\frac{(-1)^{j-1}}{4j-1}
\binom{4j-1}{j-1}.
}
\end{equation}

For $k=1$, using
\begin{equation}
[y^2](y\cot y)^j=-\frac{j}{3},
\end{equation}
we obtain
\begin{equation}
{
[x^2]p_j(x)
=
\frac{(-1)^j}{12}
\binom{4j-4}{j-3},
\qquad j\geq 3.
}
\end{equation}

Similarly,
\begin{equation}
[y^4](y\cot y)^j
=
\frac{j(5j-7)}{90},
\end{equation}
which gives
\begin{equation}
{
[x^4]p_j(x)
=
\frac{(-1)^{j-1}(5j-7)}{1440}
\binom{4j-6}{j-5},
\qquad j\geq 5.
}
\end{equation}

Finally,
\begin{equation}
[y^6](y\cot y)^j
=
-\frac{j(35j^2-147j+124)}{5670},
\end{equation}
and hence
\begin{equation}
{
[x^6]p_j(x)
=
\frac{(-1)^j(35j^2-147j+124)}{362880}
\binom{4j-8}{j-7},
\qquad j\geq 7.
}
\end{equation}

The first columns can therefore be summarized as
\begin{align}
[x^0]p_j(x)
&=
\frac{(-1)^{j-1}}{j}
\binom{4j-2}{j-1},
\\[1mm]
[x^2]p_j(x)
&=
\frac{(-1)^j}{12}
\binom{4j-4}{j-3},
\\[1mm]
[x^4]p_j(x)
&=
\frac{(-1)^{j-1}(5j-7)}{1440}
\binom{4j-6}{j-5},
\\[1mm]
[x^6]p_j(x)
&=
\frac{(-1)^j(35j^2-147j+124)}{362880}
\binom{4j-8}{j-7}.
\end{align}

More generally, the coefficient of $x^{2k}$ has the structure
\begin{equation}
[x^{2k}]p_j(x)
=
(-1)^{j+k-1}
R_{k-1}(j)
\binom{4j-2k-2}{j-2k-1},
\end{equation}
where $R_{k-1}(j)$ ({for $k>0$}) is a polynomial in $j$ of degree ($k-1$), including an overall rational normalization.

It is also possible to resum all powers of (q) at each fixed order in (x). Let $r=r(q)$ be the solution of
\begin{equation}
q=r(1+r)^3,
\qquad
r(0)=0.
\label{eq:r-algebraic}
\end{equation}
Its expansion is
\begin{equation}
r(q)
=
q-3q^2+15q^3-91q^4+612q^5+\mathcal{O}(q^6),
\end{equation}
and it resums the complete $x^0$ contribution of (\ref{eq:implicit-s}). It can also be written as
\begin{equation}
r(q)
=
q \, _4F_3\left(
\begin{matrix}
1,\frac{3}{4},\frac{5}{4},\frac{3}{2}\\
2,\frac{4}{3},\frac{5}{3}
\end{matrix};
-\frac{256q}{27}
\right),
\end{equation}
with $F_3$ the hypergeometric function.

In terms of $r(q)$, the solution of Eq.~\eqref{eq:implicit-s} can also be expanded in $x$, while keeping the dependence on $q$ resummed
\begin{align}
s(q,x)
={}&r
-\frac{x^2r^3(1+r)}
{12(1+4r)}
+
\frac{x^4r^5(1+r)(3+9r-2r^2)}
{240(1+4r)^3}
\nonumber\\
&+
\frac{x^6r^7(1+r)}
{60480(1+4r)^5}
\left(
1112r^4+2708r^3+18r^2-648r-135
\right)
+\mathcal{O}(x^8),
\end{align}
where $r=r(q)$.

Therefore, the solution can be expressed as
\begin{equation}
{
u_1(\beta)
=
\frac{n\pi}{L}
+
\zeta
\left(
1+
s\left(
\frac{\beta^2}{L\zeta^4},
L\zeta
\right)
\right),
}
\end{equation}
which coincides with \eqref{u335u1expansion}. 
At every fixed order in $x^2=(L\zeta)^2$, all powers of $\beta$ are thus resummed.

\subsection{Exact solution for arbitrary $\gamma$}

We now consider the Bethe equation for arbitrary $\gamma$. It can be rewritten as
\begin{equation}
\beta^2
=
\frac{
2(Lu_1-\pi n)
\left[(Lu_1-\pi n)^2-3\gamma L^2\right]
\tan\left(\frac{Lu_1}{2}\right)
}{L^3}.
\label{eq:BE-gamma}
\end{equation}
We introduce again as we did earlier
\begin{equation}
u_1
=
\frac{n\pi}{L}+\zeta+t(\beta),
\qquad t(0)=0.
\end{equation}
Using the zeroth-order condition ({\it i.e.} $\beta^0$), which remains the same as in the $\gamma=0$ case even if $\gamma$ is generic, that is
\begin{equation}
n\pi+L\zeta=2m\pi,
\qquad m\in\mathbb{Z},
\end{equation}
we obtain
\begin{equation}
\beta^2
=
2(\zeta+t)
\left[(\zeta+t)^2-3\gamma\right]
\tan\left(\frac{Lt}{2}\right).
\label{eq:t-gamma}
\end{equation}

We define
\begin{equation}
F_\gamma(t)
=
2(\zeta+t)
\left[(\zeta+t)^2-3\gamma\right]
\tan\left(\frac{Lt}{2}\right).
\end{equation}
The solution on the branch satisfying $t(0)=0$ is therefore
\begin{equation}
{
u_1(\beta)
=
\frac{n\pi}{L}
+\zeta
+F_\gamma^{-1}(\beta^2)
}.
\label{eq:exact-gamma}
\end{equation}

The inverse function can be expanded around $\beta=0$ using the
Lagrange--B\"urmann inversion formula. We write
\begin{equation}
u_1(\beta)
=
\frac{n\pi}{L}
+\zeta
+\sum_{k=1}^{\infty}c_k\,\beta^{2k}.
\end{equation}
The coefficients are given exactly by
\begin{equation}
{
c_k
=
\frac{1}{k}
\left[t^{k-1}\right]
\left[
\frac{t}{
2(\zeta+t)
\left((\zeta+t)^2-3\gamma\right)
\tan\left(\frac{Lt}{2}\right)
}
\right]^k
}.
\label{eq:ck-gamma}
\end{equation}
Here, $[t^m]f(t)$ denotes the coefficient of $t^m$ in the Taylor
expansion of $f(t)$.

It is convenient to introduce the dimensionless variables
\begin{equation}
x=L\zeta,
\qquad
g=\frac{3\gamma}{\zeta^2}.
\end{equation}
The coefficients can then be written as
\begin{equation}
{
c_k
=
\frac{\mathcal{P}_k(x,g)}
{L^k\zeta^{4k-1}(1-g)^{2k-1}}
},
\label{eq:ck-polynomial-gamma}
\end{equation}
where
\begin{equation}
\mathcal{P}_k(x,g)
=
\frac{(1-g)^{2k-1}}{k}
\left[z^{k-1}\right]
\left[
\frac{
\frac{xz}{2}\cot\left(\frac{xz}{2}\right)
}{
(1+z)\left((1+z)^2-g\right)
}
\right]^k.
\label{eq:Pk-gamma}
\end{equation}
The function $\mathcal{P}_k(x,g)$ is a polynomial in $x^2$ and $g$,
with
\begin{equation}
\deg_{x^2}\mathcal{P}_k
=
\left\lfloor\frac{k-1}{2}\right\rfloor,
\qquad
\deg_g\mathcal{P}_k=k-1.
\end{equation}

The first polynomials are
\begin{align}
\mathcal{P}_1(x,g)
&=1,
\\
\mathcal{P}_2(x,g)
&=g-3,
\\
\mathcal{P}_3(x,g)
&=
15-9g+2g^2
-\frac{(1-g)^2}{12}x^2,
\\
\mathcal{P}_4(x,g)
&=
-91+77g-31g^2+5g^3
+\frac{(3-g)(1-g)^2}{3}x^2,
\\
\mathcal{P}_5(x,g)
&=
612-663g+372g^2-111g^3+14g^4
\nonumber\\
&\quad
-\frac{5}{4}(1-g)^2
\left(g^2-5g+8\right)x^2
+\frac{(1-g)^4}{80}x^4.
\end{align}
For $g=0$, these expressions reduce to the polynomials obtained in
the $\gamma=0$ case.

Therefore, the expansion begins as
\begin{align}
u_1(\beta)
={}&
\frac{n\pi}{L}+\zeta
+\frac{\beta^2}{L\zeta^3(1-g)}
+\frac{g-3}{L^2\zeta^7(1-g)^3}\,\beta^4
\nonumber\\
&+
\frac{
15-9g+2g^2-\frac{(1-g)^2}{12}x^2
}{
L^3\zeta^{11}(1-g)^5
}\,\beta^6
\nonumber\\
&+
\frac{
-91+77g-31g^2+5g^3
+\frac{(3-g)(1-g)^2}{3}x^2
}{
L^4\zeta^{15}(1-g)^7
}\,\beta^8
+\mathcal{O}(\beta^{10}),
\end{align}
where $x=L\zeta$ and $g=3\gamma/\zeta^2$.

It is also useful to reorganize the solution as an expansion in
$x=L\zeta$. We introduce
\begin{equation}
q
=
\frac{\beta^2}{L\zeta^4(1-g)},
\qquad
u_1(\beta)
=
\frac{n\pi}{L}+\zeta+\zeta\,s(q,x,g).
\end{equation}
Equation~\eqref{eq:t-gamma} becomes
\begin{equation}
q
=
\frac{2}{x(1-g)}
(1+s)\left[(1+s)^2-g\right]
\tan\left(\frac{xs}{2}\right).
\label{eq:implicit-s-gamma}
\end{equation}

We expand
\begin{equation}
s(q,x,g)
=
\sum_{j=1}^{\infty}p_j(x,g)q^j,
\end{equation}
where
\begin{equation}
p_j(x,g)
=
\frac{1}{j}
\left[z^{j-1}\right]
\left[
\frac{
(1-g)\frac{xz}{2}
\cot\left(\frac{xz}{2}\right)
}{
(1+z)\left((1+z)^2-g\right)
}
\right]^j.
\label{eq:pj-gamma}
\end{equation}
The relation between the two sets of polynomials is
\begin{equation}
p_j(x,g)
=
\frac{\mathcal{P}_j(x,g)}{(1-g)^{j-1}}.
\end{equation}

To isolate the coefficient of a fixed power of $x$, we define
\begin{equation}
A_{j,k}
=
\left[y^{2k}\right](y\cot y)^j
\end{equation}
and
\begin{equation}
C_{j,r}(g)
=
\left[z^r\right]
\left[
\frac{1-g}{
(1+z)\left((1+z)^2-g\right)
}
\right]^j.
\end{equation}
Then
\begin{equation}
{
[x^{2k}]p_j(x,g)
=
\frac{A_{j,k}}{j\,4^k}
C_{j,j-1-2k}(g)
}.
\label{eq:fixed-x-gamma}
\end{equation}
This formula is valid for
\begin{equation}
0\leq k\leq
\left\lfloor\frac{j-1}{2}\right\rfloor.
\end{equation}

For $g=0$, one has
\begin{equation}
C_{j,r}(0)
=
[z^r](1+z)^{-3j}
=
(-1)^r
\binom{3j+r-1}{r},
\end{equation}
and eq.~\eqref{eq:fixed-x-gamma} reduces to the binomial formula
obtained in the $\gamma=0$ case.

It is also possible to resum all powers of $q$ at each fixed order
in $x$. At order $x^0$, let $r=r(q,g)$ be the solution of
\begin{equation}
q
=
\frac{
r(1+r)\left[(1+r)^2-g\right]
}{1-g},
\qquad
r(0,g)=0.
\label{eq:r-gamma}
\end{equation}
Equivalently, $r(q,g)$ is the perturbative root of the quartic equation
\begin{equation}
r^4+3r^3+(3-g)r^2+(1-g)r-(1-g)q=0.
\end{equation}

Writing
\begin{equation}
s(q,x,g)
=
r(q,g)+x^2s_1(q,g)+\mathcal{O}(x^4),
\end{equation}
we obtain
\begin{equation}
s_1(q,g)
=
-
\frac{
r^3(1+r)\left[(1+r)^2-g\right]
}{
12\left[
4r^3+9r^2+6r+1-g(2r+1)
\right]
},
\end{equation}
where $r=r(q,g)$. This is because we expand the original definition of $q$ in series of $x$ like so
\begin{eqnarray}
q = \frac{s(1+s)\Big((s+1)^2-g\Big)}{1-g} + x^2 \, \frac{s^3(1+s)\Big((s+1)^2-g\Big)}{12(1-g)}+... \label{exp}
\end{eqnarray}
At order $x^0$, we recover the quartic equation above. We write this as
\begin{eqnarray}
    q = \xi(s) + x^2 \, \tau(s)+...,
\end{eqnarray}
having introduced two new symbols for the functions in the expansion (\ref{exp}). By indicating as $s_0=r$, iterating the expansion gives
\begin{eqnarray}
    q = \xi(s_0) + (s-s_0) \, \xi'(s_0) + x^2 \, \tau(s_0) + ...,
\end{eqnarray}
having kept only the first non-trivial order. Since $q = \xi(s_0)$ by construction, we  find that it must be
\begin{eqnarray}
    (s-s_0) \, \xi'(s_0) + x^2 \, \tau(s_0) = 0
\end{eqnarray}
hence
\begin{eqnarray}
    s_1 = - \frac{\tau(s_0)}{\xi'(s_0)}. 
\end{eqnarray}
Therefore,
\begin{align}
s(q,x,g)
={}&
r(q,g)
\nonumber\\
&-
x^2
\frac{
r(q,g)^3\left[1+r(q,g)\right]
\left(\left[1+r(q,g)\right]^2-g\right)
}{
12\left[
4r(q,g)^3+9r(q,g)^2+6r(q,g)+1
-g\left(2r(q,g)+1\right)
\right]
}
+\mathcal{O}(x^4).
\end{align}

Finally, the solution can be written as
\begin{equation}
{
u_1(\beta)
=
\frac{n\pi}{L}
+
\zeta
\left[
1+
s\left(
\frac{\beta^2}{L\zeta^4(1-g)},
L\zeta,
g
\right)
\right],
\qquad
g=\frac{3\gamma}{\zeta^2}.
}
\end{equation}
At each fixed order in $x^2=(L\zeta)^2$, all powers of $\beta$ are
resummed in terms of the same algebraic function $r(q,g)$.

The above expansion assumes
\begin{equation}
\zeta^2\neq3\gamma,
\end{equation}
or equivalently $g\neq1$. At the special point $g=1$, the linear term
of $F_\gamma(t)$ vanishes and the usual expansion in powers of
$\beta^2$ must be modified.

\section{Conclusions}
\label{conclusions}
In this paper, we have studied the higher conserved charges of the quantised Kadomtsev-Petviashvili equation, with particular attention to the first higher charge (beyond the basic Noether charges associated with mass, momentum and Hamiltonian which were already displayed in \cite{Kozlowski:2016too}). The explicit construction of even the first higher charge is non-trivial and traditionally non-standard, in the sense that, even in simpler models such as the non-linear Schr\"odinger equation, normal-ordering the classical expression when transferring to the quantum version is typically not enough, \cite{nls}. We decided to brute-force construct the quantum operator associated with the first higher charge and check that it commutes with the lower charges. We were aided in this endeavour by a judicious use of the Galilei boost operator. To conclude the proof, we need to introduce a regulator of a certain  ultraviolet divergence - we prove that the result in the end does not depend on the particular choice of regulator, under fairly general assumptions.

A natural direction for future work is to investigate whether there exists 
a quantum boost operator, see \cite{links2001ladder},  connecting the charges $H_2$ and $H_3$, which could 
then be used to construct the higher conserved charges of the theory and, 
possibly, to detect integrability directly in quantum field theory - see also \cite{Fiora}.

We then presented a complete solution of the two-particle Bethe ansatz in finite volume, displaying an iterative solution expressed in powers of one of the couplings, obtained with the help of modern AI tools.

In this regard, there are several directions that can be pursued. 

It would be interesting to investigate whether an appropriate large-coupling limit reproduces the nonlinear Schr\"odinger model.

Another interesting direction is to pursue the approach of generalised hydrodynamics \cite{Doyon} in the case of the quantised KP equation as we have been analysing in this paper, with particular attention to the issue of the higher conserved charges.  Recent developments applying generalised hydrodynamics to classical Boussinesq and KP soliton gases make this a particularly natural direction to explore, see \cite{doyonhydro}.

It is amusing to notice the appearance of a cubic type of $S$-matrix of the same type as the one relevant for quantum KP in the definition of the definition of the affine Yangian, see for instance  \cite{Gaberdiel:2017dbk} - formula (2.4). Precisely the same condition of vanishing of the total coefficient of the quadratic power at the numerator and denominator is observed in either cases.

Numerical studies of the KP solitons - see for instance \cite{SB} - can elucidate further properties of the theory and are left for an interesting future prospect.

\section{Appendix A}

\label{AppendixcommutatorsH0H1H2}

\subsection{$[H_0,H_1]$}
Using \eqref{H0PsiandH0Psidag} we get
\begin{align}
&    [H_0,\partial_x\Psi_m(x)]
=
-m\,\partial_x\Psi_m(x), &[H_0,\partial_x\Psi^\dagger_m(x)]
=
-m\,\partial_x\Psi^\dagger_m(x).
\end{align}

Therefore,
\begin{align}
\begin{aligned}
[H_0,H_1]
&=
-i\sum_{m\geq 1}
\int dx\,
[H_0,\Psi_m^\dagger(x)\partial_x\Psi_m(x)]
\\
&=
-i\sum_{m\geq 1}
\int dx\,
\left(
m\,\Psi_m^\dagger(x)\partial_x\Psi_m(x)
-
m\,\Psi_m^\dagger(x)\partial_x\Psi_m(x)
\right)
=0.
\end{aligned}
\end{align}

Hence,
\begin{align}
{
[H_0,H_1]=0.
}
\end{align}

\subsection{$[H_0,H_2]$}

Using \eqref{H0PsiandH0Psidag}, we also have
\begin{align}
&[H_0,\partial_x^2\Psi_m(x)]
=
-m\,\partial_x^2\Psi_m(x),
 &[H_0,\partial_x^2\Psi^\dagger_m(x)]
=
m\,\partial_x^2\Psi^\dagger_m(x).
\label{H0PsiPsidag}
\end{align}

For the quadratic kinetic term,
\begin{align}
\begin{aligned}
[H_0,\Psi_m^\dagger(x)\partial_x^2\Psi_m(x)]
&=
m\,\Psi_m^\dagger(x)\partial_x^2\Psi_m(x)
-
m\,\Psi_m^\dagger(x)\partial_x^2\Psi_m(x)=0.
\end{aligned}
\end{align}

Similarly,
\begin{align}
[H_0,\Psi_m^\dagger(x)\Psi_m(x)]=0.    
\end{align}

Now consider the joining cubic term \eqref{Jjoining}, then
\begin{align}    
[H_0,J_{m_1,m_2}(x)]
=
\left(M-m_1-m_2\right)
J_{m_1,m_2}(x)
=0.\label{commutH0J}
\end{align}

Similarly, for the splitting cubic term \eqref{Jsplitting}, we get
\begin{align}
[H_0,J_{m_1,m_2}^\dagger(x)]
=
\left(m_1+m_2-M\right)
J_{m_1,m_2}^\dagger(x)
=0.\label{commutH0Jdag}
\end{align}

Therefore every term in \(H_2\) commutes with \(H_0\), and hence
\begin{align}
{
[H_0,H_2]=0.}    
\end{align}

\subsection{$[H_1,H_2]$}

The charge \(H_1\) generates translations in \(x\), namely
\begin{align}
[H_1,\Psi_m(x)]
=
i\,\partial_x\Psi_m(x),
\qquad
[H_1,\Psi_m^\dagger(x)]
=
i\,\partial_x\Psi_m^\dagger(x).
    \end{align}

Therefore, for any local density \(\mathcal O(x)\) without explicit \(x\)-dependence (for example $x \Psi(x)$ won't be good),
\begin{align}
[H_1,\mathcal O(x)]
=
i\,\partial_x\mathcal O(x).    
\end{align}

For example, for the kinetic density we have
\begin{align}
\begin{aligned}
[H_1,\Psi_m^\dagger(x)\partial_x^2\Psi_m(x)]
&=
i(\partial_x\Psi_m^\dagger(x))\partial_x^2\Psi_m(x)
+
i\Psi_m^\dagger(x)\partial_x^3\Psi_m(x)
\\
&=
i\partial_x
\left(
\Psi_m^\dagger(x)\partial_x^2\Psi_m(x)
\right).
\end{aligned}    
\end{align}

Similarly,
\[
[H_1,\Psi_m^\dagger(x)\Psi_m(x)]
=
i\partial_x
\left(
\Psi_m^\dagger(x)\Psi_m(x)
\right).
\]

For the joining cubic term, using the definition of $J_{m_1,m_2}$ of \eqref{Jjoining},
we obtain
\begin{align}
\begin{aligned}
[H_1,J_{m_1,m_2}(x)]
&=
i(\partial_x\Psi_M^\dagger(x))\Psi_{m_1}(x)\Psi_{m_2}(x)
+i\Psi_M^\dagger(x)(\partial_x\Psi_{m_1}(x))\Psi_{m_2}(x)
\\
&\quad
+i\Psi_M^\dagger(x)\Psi_{m_1}(x)(\partial_x\Psi_{m_2}(x))
=
i\partial_x J_{m_1,m_2}(x).
\end{aligned}
\end{align}

The same argument applies to the splitting term \eqref{Jsplitting},
\begin{align}
[H_1,J_{m_1,m_2}^\dagger(x)]
=
i\partial_x J_{m_1,m_2}^\dagger(x).    
\end{align}

Hence, the full density \(h_2(x)\) satisfies
\begin{align}
[H_1,h_2(x)]
=
i\,\partial_x h_2(x).    
\end{align}

Therefore,
\begin{align}
[H_1,H_2]
=
\int dx\,[H_1,h_2(x)]
=
i\int dx\,\partial_x h_2(x).    
\end{align}

Assuming periodic boundary conditions, or fields decaying sufficiently fast at infinity,
\begin{align}
\int dx\,\partial_x h_2(x)=0.
\end{align}

Thus,
\begin{align}
{
[H_1,H_2]=0.
}    
\end{align}

\section*{Acknowledgements}The authors gratefully acknowledge significant assistance from OpenAI's ChatGPT (GPT-5.6 Pro) in the derivation, verification, and presentation of several results. The authors carefully reviewed the AI-assisted arguments, proofs, and calculations, revised their final exposition, and take full responsibility for the contents of the manuscript. A.T. also thanks the student Wookyung Kim for having worked on a Master project on the theme of the quantum KP equation. C.P. acknowledge funding from the European Union HORIZON-CL4-2022 QUANTUM-02-SGA through PASQuanS2.1 (Grant Agreement No. 101113690), European Research Council (ERC) through Advanced grant QUEST (Grant Agreement No. 101096208) and ARIS, the Slovenian Research and Innovation Agency, under the Seal of Excellence scheme through DeCiQuC.

\end{document}